\documentstyle[preprint,aps]{revtex}
\begin{document}
%\draft
%\preprint{MAD/TH/92-1}
\title{%
Two-loop renormalization constants and high energy $2\rightarrow2$\\
scattering amplitudes in the Higgs sector of the standard model}
\author{Peter N. Maher\cite{pm}, Loyal
Durand\cite{ld}, and Kurt Riesselmann\cite{kr}}
\address{%
Department of Physics, University of
Wisconsin--Madison\\
Madison, Wisconsin 53706}
\maketitle
\begin{abstract}
We calculate the complete matrix of two-body scattering amplitudes
for the scattering of longitudinally polarized gauge bosons
$W_L^\pm$, $Z_L$ and Higgs bosons to two loops in the high-energy,
heavy-Higgs limit $\sqrt{s}\gg M_H\gg M_W$. Use of the Goldstone
boson equivalence theorem reduces the problem to one involving
only the scalar fields $w^\pm$, $z$ (the Goldstone bosons of the
original theory) and the Higgs boson. Renormalization of the
scattering amplitudes requires the calculation of the self-energy
functions $\Pi _i^0(M_i^2)$, the renormalization constants $Z_i$,
and the bare quartic Higgs coupling $\lambda _0$ to two loops. The
results will be useful in other calculations. To facilitate the
calculations, we introduce a powerful new technique for evaluating
integrals over Feynman parameters in dimensional regularization
which is based on a Barnes' type representation of the binomial
expansion. We also collect some useful integrals which extend the
tables given by Devoto and Duke.
\end{abstract}


\narrowtext
\section{INTRODUCTION}

It was pointed out some time ago by Dicus and Mathur \cite{dw1} and
Lee, Quigg, and Thacker \cite{lqt} that the presently unknown mass
$M_H$ of the Higgs boson could be bounded above in a
weakly-interacting standard model by using the constraint on the
magnitude of partial-wave scattering amplitudes implied by
unitarity. This was examined in detail in \cite{lqt}, where it was
shown that the only real constraints from two-body scattering arise
from processes which involve only longitudinally polarized gauge
bosons $W_L^\pm,Z_L$ and Higgs bosons $H$. These processes are
enhanced by factors of $M_H^2/M_W^2\propto\lambda /g^2$ relative to
those which involve transversely polarized gauge bosons and the
small gauge couplings $g$. Here $\lambda $ is the quartic coupling
for the Higgs boson, and is related to $M_H$ and the vacuum
expectation value $v=\left( \sqrt{2}G_F \right)^{-1/2}=246$ GeV by
$\lambda =M_H^2/2v^2$. When $\lambda $ (or $M_H$) is large enough,
the interactions of the Higgs sector of the standard model become
strong, and the unitarity constraints on $\lambda $ or $M_H$ are to
be interpreted in terms of a transition from weak to strong
coupling\cite{lqt,velt,cg}, rather than as upper bounds on either.

The original tree-level analysis of scattering in the neutral
channels $W_L^+W_L^-$, $Z_LZ_L$, $HH$, and $Z_LH$ given in
\cite{lqt} has been extended to one loop by a number of authors
\cite{pass1,dw1,dw2,mvw,pass2,djl1,lend,djl2}, and sharpened at tree
level \cite{mvw,vay} and at one loop \cite{djl1,djl2} using
renormalization-group arguments. The most restrictive analysis was
that in \cite{djl1,djl2} where it was shown that strong coupling
sets in for $\lambda \agt2.2$--2.5, a value substantially below the
tree-level bound $\lambda \approx8\pi /3\approx8.38$ (or $M_H=1.007$
TeV) derived by Dicus and Mathur \cite{dm} and Lee, Quigg, and
Thacker \cite{lqt}. It is desirable to extend the analysis to two
loops. This is important as a check that the conclusions of
\cite{djl1,djl2} are in fact correct, and that an increase of one
order in perturbation theory does not loosen the (unexpectedly low)
upper bound on $\lambda $  in a weakly interacting
theory.

The present paper is devoted to the calculation of the two-body
scattering amplitudes for $W_L^\pm,Z_L,H$ scattering to two loops in
the high-energy, heavy-Higgs limit $M_W\ll M_H\ll\sqrt{s}$. Our
results are used in the following paper \cite{dmr} to re\,examine
the limits on $\lambda $ and $M_H$
to two loops in a weakly interacting theory.

The organization of this paper is as follows. In Sec.\ \ref{general},
we discuss the general formulation of the problem, its reduction
through the use of the Goldstone boson equivalence theorem to a
problem involving only the scalar sector of the unbroken standard
model, and the renormalization of the scalar Lagrangian. In Sec.\
\ref{calculation}, we discuss the calculation of the self-energies
and wave function renormalization constants to two loops. The most
difficult parts of the calculation are in the extraction of the
finite parts of the two-loop terms. We sketch the methods which we
found useful for various diagrams, but give only minimal details.
The results are summarized by diagram in Appendix \ref{appa}, while
some technical points of the methods used are discussed in
Appendices B, C and \mbox{D}. In particular, we illustrate in
Appendix B a powerful new method for evaluating integrals over
Feynman parameters in dimensional regularization which is based on a
Barnes'-type representation of the binomial expansion. We sketch the
use of Kotikov's method of differential equations to evaluate
certain two-loop graphs in Appendix C, and collect a number of
useful integrals in Appendix D.

The complete self-energy counterterms and the wave function and
coupling renormalizations are collected in Sec.\ \ref{calculation}D.
These and the renormalized two-loop scattering amplitudes given in
Sec.\ IV are the principal results of the paper, and constitute the
input for \cite{dmr}. The counterterms and renormalization
constants are also directly relevant to calculations of
$W_L^\pm,Z_L$ scattering at ``low" energies for very massive Higgs
bosons, that is, for $M_W\ll\sqrt{s}\alt M_H$.

\section{GENERAL FORMULATION OF THE PROBLEM}\label{general}
\subsection{Framework}\label{background}

We will be concerned in the following sections with the
calculation to two loops (O$(\lambda^3)$) of the complete set of
two-body scattering amplitudes for high-energy scattering in the
coupled neutral channels $W_L^+W_L^-$, $Z_LZ_L$, $HH$, $HZ_L$.
The calculation can be simplified
enormously in the limit we will consider, $\sqrt{s}\gg M_H\gg
M_W$, by the use of the Goldstone boson equivalence theorem
\cite{lqt,cg,corn,bs,hvelt,he}. This theorem states that the
scattering amplitudes for $n$ longitudinally-polarized vector
bosons $W_L^\pm$,$Z_L$ and any number of other external particles
are related to the corresponding scattering amplitudes for the
scalar bosons $w^\pm$, $z$ to which $W_L^\pm$, $Z_L$ reduce for
vanishing gauge couplings $g,g'$ by
\FL
\begin{equation}
T(W_L^\pm,Z_L,H,\ldots)=(iC)^nT(w^\pm,z,H,\ldots)+
\mbox{O}(M_W/\sqrt{s}).
\label{eq1}
\end{equation}

The constant $C$ depends on the normalization scheme used in the
calculation \cite{bs,he,fixing},
\begin{equation}
C=\frac{M_W^0}{M_W}\,\left(\frac{Z_W}{Z_w}\right)^{1/2}
\left[1+{\rm O}(g^2)\right],\label{eq2}
\end{equation}
where the $Z$'s are the wave function renormalization constraints
for the physical fields $W^\pm$ and the scalar fields $w^\pm$. $C$
is equal to unity for $g,g'\rightarrow0$ in schemes in which the
renormalizations are defined at mass scales $m\ll M_H$. We will
renormalize the $w^\pm$, $z$ fields at $m=0$, a choice which
corresponds to massless Goldstone bosons and is consistent with
the assumption that $M_W\ll M_H$. Then \cite{bs,he},
\begin{equation}
C=1+{\rm O}(g^2).\label{eq3}
\end{equation}
Since we are interested in upper limits on $\lambda$ or $M_H$, the
limit $M_W^2/M_H^2\propto
g^2/\lambda\ll1$ required for the validity of
the theorem is natural for our purposes. We will henceforth set
the gauge couplings to zero, and work entirely with the scalar
fields.

\subsection{Lagrangian and renormalization}

The original Lagrangian for the symmetry-breaking sector of the
standard model,
\begin{equation}
{\cal L}_0=\frac{1}{2}(D_\mu\Phi)^\dagger(D^\mu\Phi)-
\frac{\lambda}{4}(\Phi^\dagger\Phi-u^2)^2
\label{eq4}
\end{equation}
with $D^\mu = \partial^\mu + ig{\bf W}^\mu \cdot {\bf T}
+\frac{1}{2} g' B^\mu$ and
\begin{equation}
\Phi=\left(\begin{array}{c@{+}c}
w_1 & iw_2\\
h & iw_3
\end{array}\right),
\label{eq5}
\end{equation}
reduces for $g,g'\rightarrow0$ to an SO(4)-symmetric Lagrangian
involving only the scalar fields $h$ and ${\bf w}=(w_1,w_2,w_3)$.
The field $h$ will be taken as usual as the component of $\Phi$
which acquires a vacuum expectation value,
$u$ at tree level and $v$ in the full theory.
We
will write $h$ as $h=v+H$ where $\langle\Omega|H|\Omega\rangle=0$
with respect to the physical vacuum $|\Omega\rangle$. Adding all
possible SO(4)-symmetric counterterms and rewriting
the potential in ${\cal L}_0$ in terms of $v$ rather than $u$,
we obtain the
complete Lagrangian for $g=g'=0$,
\FL
\begin{eqnarray}
{\cal L} &=& \case{1}/{2}(\partial_\mu\Phi)^\dagger
(\partial^\mu\Phi)+\case{1}/{2}A(\partial_\mu\Phi)^\dagger
(\partial^\mu\Phi)\nonumber\\
&&-\left(\case{1}/{4}\lambda(\Phi^2-v^2)^2+
\case{1}/{2}\lambda(v^2-u^2)(\Phi^2-v^2)+
\case{1}/{2}\delta
m^2\Phi^2+\case{1}/{4}\delta\lambda\Phi^4\right). \label{eq6}
\end{eqnarray}
$\Phi$ is now to be written in terms of $\bf w$, $H$, and $v$.

The SO(4) symmetry of the unbroken scalar theory and the condition
$\langle\Omega|H|\Omega\rangle=0$ impose  constraints on
$u^2$, $A$, $\delta m^2$, and $\delta\lambda$.
For example, Higgs tadpole diagrams shift the vacuum  expectation
value and contribute apparent mass terms for the
Goldstone bosons which must be absent in the final theory.
Taylor
\cite{taylor} has outlined a procedure which sums the tadpole
diagrams to all orders, and enforces the vanishing of
$\langle\Omega|H|\Omega\rangle$.
The same condition guarantees the vanishing of the masses of the
Goldstone bosons $\bf w$, and allows the tree-level vacuum
expectation value $u$ to be eliminated in terms of $v$, the
self-energy function $\Pi_w(0)$, and the remaining parameters
\cite{maher}. The parameter $A$
appears in the kinetic terms for both the $w$ and $H$ fields
because of the
underlying symmetry in the problem.  It
is related to
the wave function renormalization constant
$Z_w$ by $Z_w=1+A$.
Finally, $\delta m^2$ and $\delta\lambda$ are related through the
constraint $M_H^2=2\lambda v^2$,
which is taken to be exact, that is, to hold to all orders in
perturbation theory, with $M_H$ the physical mass
of the Higgs boson defined by the condition ${\rm Re}\,
(i\Delta_H^{-1}(M_H^2))=0$ \cite{maher}. In enforcing this
constraint, we are choosing a renormalization scheme and
definitions of $M_H$ and $\lambda$ which are consistent with those
introduced by Sirlin \cite{sirlin} and by Sirlin and Zucchini
\cite{sz} for the full theory,
with $g,g'\not=0$.
The result of this analysis
\cite{maher} is a modified Lagrangian
\begin{eqnarray}
{\cal L}&=&\case{1}/{2}\partial_\mu{\bf w}_0\cdot\partial^\mu
{\bf w}_0+\case{1}/{2}\partial_\mu H_0\,\partial^\mu H_0
-\case{1}/{2}\left(M_H^2-{\rm Re}\,\Pi^0_H(M_H^2)\right)H_0^2
+\case{1}/{2}\Pi^0_w(0){\bf w}_0^2\nonumber\\
&&-\case{1}/{4}\lambda\left(%
1-\frac{{\rm Re}\,\Pi^0_H(M_H^2)-\Pi^0_w(0)}{M_H^2}\right)
\left[\frac{1}{Z_w}({\bf w}_0^2+H_0^2)^2
+\frac{4v}{Z_w^{1/2}}({\bf w}_0^2+H_0^2)\right].
\label{eq7}
\end{eqnarray}
We have introduced bare fields ${\bf w}_0=Z_H^{1/2}{\bf w}$ and
$H_0=Z_w^{1/2}H$ to keep the kinetic terms and free propagators in
standard form \cite{physical}. $\Pi^0_w$ and $\Pi^0_H$ are the
self-energy functions for the bare fields
and are to be calculated using the Lagrangian in Eq.\ (7). We will
later drop the superscript on $\Pi^0$ for simplicity in denoting
the one- and two-loop contributions to these functions,
$\Pi^0=\Pi^{(1)}+\Pi^{(2)}+\cdots$.
The scattering amplitudes
calculated in terms of ${\bf w}_0$ and $H_0$  require the usual wave
function renormalization, so must be multiplied by a factor
$Z_i^{1/2}$ for each external line.

The form of the new Lagrangian clearly ensures that the Goldstone
bosons $\bf w$ are massless and that the Higgs boson has
the physical
mass
$M_H$. The constraint $M_H^2=2\lambda v^2$ can be used to
eliminate either $M_H^2$ or $\lambda$ given the physical value of
$v$. However, it will be most convenient to keep ${\cal L}$ as
written, and only use the constraint as needed. Finally, in using
this form of the Lagrangian, all Higgs tadpole graphs should be
omitted since the shift in the vacuum expectation value from $u$
to $v$ has been suppressed following Taylor \cite{taylor}.

The SO(4) symmetry of the original Lagrangian is broken by the
mass terms and the trilinear couplings in Eq.\ (\ref{eq7}), but
remains in the quartic couplings. There is still a residual SO(3)
symmetry for the $\bf w$ fields, so the self-energy functions and
renormalization constants for the charged and neutral particles
$w^\pm=(w_1\mp iw_2)/\sqrt{2}$ and $z=w_3$ are identical,
\begin{equation}
Z_z^{-1}=Z_w^{-1}=1-\frac{d}{dp^2}\Pi^0_w(p^2)|_{p^2=0}.
\label{eq8}
\end{equation}
However, the self-energy function and wave function
renormalization constant for the Higgs boson are different, and
the subtraction point is at $p^2=M_H^2$ rather than $p^2=0$,
\begin{equation}
Z_H^{-1}=1-\frac{d}{dp^2}
{\rm Re}\Pi^0_H(p^2)|_{p^2=M_H^2}.
\label{eq9}
\end{equation}

The SO(4) symmetry of the theory is partially restored in
scattering processes in the high energy limit $\sqrt{s}\gg M_H$.
In particular, scattering graphs which involve the dimension-three
trilinear couplings in Eq.\ (\ref{eq7}) are suppressed by powers
of $s$, $t$, or $u$ relative to graphs which involve only the
quartic dimension-four couplings \cite{old32}, and can be
neglected for $s\gg M_H^2$ subject to some subtleties discussed
later. The only  contributions of the trilinear couplings for
$\sqrt{s}\rightarrow\infty $
occur through the renormalization constants $Z_w$, $Z_H$ which are
fixed at low energies
or, at two loops,
in the unitarity sum for the partial-wave amplitudes, through
$2\rightarrow3$ processes which have a low-energy component.
The $Z$'s and low-energy
processes  break the SO(4) symmetry even though
it is present in the high-energy interactions.

\subsection{The Feynman diagrams}

We used the programs  {\tt DIAGRAMMAR}  and {\tt
DRAW} \cite{jjohnson}
to generate the self-energy and scattering diagrams which
are needed through two loops, and to calculate their weights,
including the couplings. The results were checked by hand and by a
Wick expansion. The thirty-eight distinct diagrams for the one-
and two-loop self-energies are shown in Figs.\
\ref{fig1}--\ref{fig3} without the couplings or weights. All tadpole
graphs have been omitted as discussed above. The coupling
counterterms are not shown as these do not alter the Feynman
integrals to be calculated.

For ease of reference, the distinct topologies for the self-energy
diagrams are named {\em Scoop} (${\cal Q}$), {\em Bubble} (${\cal
B}$), {\em Triangle} (${\cal T}$), {\em Acorn} (${\cal A}$),
{\em Eye-in-the-Sky} (${\cal E}$), {\em Sunset} (${\cal S}$), and
{\em Lemon} (${\cal L}$). Individual graphs are labeled by the
number of internal Higgs lines as shown in the figures,
with extra labels $m$, $0$, and ${\cal Q}$
to distinguish otherwise-ambiguous cases. In the analytic
expressions developed in the following sections, the symbols
${\cal Q}$, ${\cal B}$, \ldots will refer specifically to the
Feynman integrals associated with the corresponding diagrams, with
the coupling constants and weights from the Wick reduction
omitted, e.g., from Fig.\ \ref{fig1}
\begin{equation}
{\cal B}_2=\int{\cal D}k\,
\frac{i}{(k^2-M_H^2)}\frac{i}{((k+q)^2-M_H^2)}
\label{eq10}
\end{equation}
for incoming momentum $q$.

The scattering diagrams which are needed through two loops are
shown in Fig.\ \ref{fig4}, with a similar labeling. The tree,
one- and two-loop graphs $V$, ${\cal B}_S$, ${\cal D}_S$ and
${\cal A}_S$ involve only the SO(4)-symmetric quartic couplings.
The complete set of
scattering
diagrams also includes, where appropriate, contributions from the
$t$ and $u$ channels. These
scattering amplitudes  are
SO(4)- and
crossing-symmetric and  increase in magnitude
 for $s\rightarrow\infty$. In contrast, the sample diagrams
shown in Fig.\ \ref{fig5}, which include trilinear couplings,
vanish for $s, |t|, |u|\rightarrow\infty$ as is easily checked.

The calculation of the self-energy terms will be discussed in the
next section. The scattering graphs will be considered in Sec.\ IV.

\section{CALCULATION OF SELF ENERGIES AND RENORMALIZATION
CONSTANTS}\label{calculation}

\subsection{The general problem}

The self-energy function $\Pi_w(p^2)$ for the Goldstone bosons
$\bf w$ is given to two loops by
\begin{eqnarray}
-i\Pi^0_w &=& \frac{-i\lambda}{Z_w}\left(
1-\frac{{\rm Re}\,\Pi_H^{(1)}
(M_H^2)-\Pi_w^{(1)}(0)}{M_H^2}\right){\cal Q}
\nonumber\\
&& +\frac{(-i\lambda v)^2}{Z_w}\left(
1-\frac{{\rm Re}\,\Pi_H^{(1)}
(M_H^2)-\Pi_w^{(1)}(0)}{M_H^2}\right)^2\,
4{\cal B}_1\nonumber\\
&& - i\lambda{\cal B}_Q[i{\rm Re}\,\Pi_H^{(1)}(M_H^2)
-i3\lambda{\cal Q}]-i\lambda(-i\lambda v)^2[18{\cal A}_{4Q}
+6{\cal A}_{2Q}+20{\cal A}_{1Q}]\nonumber\\
&& +(-i\lambda v)^2 4[(i{\rm Re}\,\Pi_H^{(1)}(M_H^2)-i3\lambda{\cal
Q}) {\cal T}_2+(i\Pi_w(0)-i\lambda{\cal Q}){\cal T}_1]\nonumber\\
&& + (-i\lambda v)^2[10{\cal S}_0+2{\cal S}_2]\nonumber\\
&& - i\lambda(-i\lambda v)^2[40{\cal A}_{10}+24{\cal A}_3+
16{\cal A}_{20}+8({\cal B}_1)^2]\nonumber\\
&& +(-i\lambda v)^4[72{\cal E}_4+24{\cal E}_2^*+16{\cal E}_2
+48{\cal L}_3+16{\cal L}_{20}],\label{eq11}
\end{eqnarray}
where ${\cal Q}$, ${\cal B}$, etc.\ refer to the Feynman integrals
which are associated with the diagrams in Figs.\
\ref{fig1}--\ref{fig3}.  We have not
given the explicit
expressions for the factors in the first two terms which
arise from the wave function and coupling renormalizations so as
to indicate their origin more clearly. These factors can be
expanded systematically using the one-loop
results from Eq.\ (11) and the following
equation and their analogs for $d\Pi^0_w/dp^2$ and
$d\Pi^0_H/dp^2$.

The
corresponding result for the Higgs boson is
\begin{eqnarray}
-i\Pi^0_H &=& 3\frac{-i\lambda}{Z_w}
\left(1-\frac{{\rm
Re}\,\Pi_H^{(1)}
(M_H^2)-\Pi_w^{(1)}(0)}{M_H^2}\right){\cal Q}\nonumber\\
&& + \frac{(-i\lambda v)^2}{Z_w}
\left(1-\frac{{\rm
Re}\,\Pi_H^{(1)}
(M_H^2)-\Pi_w^{(1)}(0)}{M_H^2}\right)^2 [18{\cal
B}_2+6{\cal B}_0]\nonumber\\
&& - i3\lambda{\cal B}_Q[i{\rm
Re}\,\Pi^{(1)}_H(M_H^2)- i3\lambda{\cal Q}]\nonumber\\
&& - i\lambda(-i\lambda v)^2[54{\cal A}_{4Q}
+18{\cal A}_{2Q}+12{\cal A}_{1Q}]\nonumber\\
&& + (-i\lambda v)^2[36(i{\rm
Re}\,\Pi_H^{(1)}(M_H^2)-i3\lambda{\cal Q}) {\cal
T}_3+12(i\Pi_w(0)-i\lambda{\cal Q}){\cal T}_0]\nonumber\\ && +
(-i\lambda)^2[6{\cal S}_1+6{\cal S}_3]-i\lambda(-i\lambda v)^2
[216{\cal A}_4+72{\cal A}_{2m}+48{\cal A}_{1m}]\nonumber\\ && -
i\lambda(-i\lambda v)^2[30({\cal B}_0)^2+36{\cal B}_0{\cal B}_2
+54({\cal B}_2)^2]\nonumber\\
&& + (-i\lambda v)^4[648{\cal E}_5 + 216{\cal E}_3 + 48{\cal E}_1
+648{\cal L}_5 + 144{\cal L}_{2m} + 24{\cal L}_1].\label{eq12}
\end{eqnarray}
We will only need these functions evaluated for $p^2=0$
($\Pi^0_w(0)$) and $p^2=M_H^2$ ($\Pi^0_H(M_H^2)$) for use in Eq.\
(\ref{eq7}).

The wave function renormalization constants $Z_w^{-1}$ and
$Z_H^{-1}$, Eqs.\ (\ref{eq8}) and (\ref{eq9}), involve similar but
more complicated expressions obtained by formally differentiating
the expressions for $\Pi^0_w$ and
$\Pi^0_H$ with respect to $p^2$, and
evaluating the results at $p^2=0$ and $p^2=M_H^2$, respectively.
We thus need expressions for all of the self-energy graphs and
their derivatives with respect to the external momentum, but only
for $p^2=0$ or $p^2=M_H^2$.

We have calculated the Feynman integrals using dimensional
regularization with the number of dimensions $D$ given by
$D=2\omega=4-2\epsilon$.
To keep the coupling $\lambda$ dimensionless in $2\omega$
dimensions, we multiply the quartic coupling in Eq.\ (\ref{eq7})
by $\mu^{4-2\omega}$ and the trilinear coupling by
$\mu^{2-2\omega}$, where $\mu$ is an arbitrary scale mass which
will drop out in the final results. For the diagrams we
need, each loop then appears with a factor $\mu^{4-2\omega}$ which
we will incorporate into the measure of the momentum integrals,
defining ${\cal D}k$ as
\begin{equation}
{\cal D}k=\mu^{4-2\omega}\frac{d^{2\omega}k}{(2\pi)^{2\omega}}.
\label{eq14}
\end{equation}
Our definitions for the momenta and particle masses in the Feynman
integrals are shown in Figs.\
\ref{fig6}, \ref{fig7}.
Ordinary Feynman parameter techniques
were used to combine propagators, and are sufficient for most of
the calculations. (However, we used other methods---hyperspherical
or Gegenbauer polynomial expansions and Kotikov's differential
equation method---to evaluate some of the {\em Lemon} (${\cal L}$)
graphs, as discussed later.) The basic integral needed after
combining propagators is \cite{thooft}
\begin{equation}
\int\,
\frac{d^{2\omega}k}{[k^2+2k\cdot q -m^2+i\epsilon]^\alpha}=
i\pi^\omega\,
\frac{\Gamma(\alpha-\omega)}{\Gamma(\alpha)}\,
\frac{e^{-i\pi\omega}}{[-q^2-m^2+i\epsilon]^{\alpha-\omega}}
\label{eq13}
\end{equation}
where $-M^2 + i\epsilon = e^{i\pi}(M^2 - i\epsilon)$.

Since a number of functions and structures appear repeatedly, we
will use some abbreviated notations. Thus, we will scale all
masses or momenta by $M_H$, and denote a scaled quantity by a
caret, e.g., $\hat s=s/M_H^2$. Each loop of a divergent graph
contributes a factor
\begin{equation}
\xi^\epsilon=(4\pi\mu^2/M_H^2)^\epsilon =
(4\pi\hat \mu^2)^\epsilon
\label{eq15}
\end{equation}
to the final result; we will generally not display this factor
except when needed. A derivative of an integral with respect to an
external momentum, $\partial(\mbox{graph})/\partial q^2$, will
often be written as $\partial(\mbox{graph})$. Finally, the first
Feynman parameter introduced, $x$, often appears in the
combination $X=x(1-x)$.

Our results for the integrals which appear in the self-energy
functions and their derivatives are given in Appendix \mbox{A}. In
the following subsections, we will simply sketch the methods used
in the calculations, with minimal detail.

\subsection{The one-loop diagrams}

The calculation of the one-loop graphs is straightforward.
However, it is necessary to keep terms of O($\epsilon$) in the
expansion of the resulting expressions for $\epsilon\rightarrow0$
in order to cover cases in which the graphs are multiplied by
terms of order $1/\epsilon$, and for articulated graphs.

\subsubsection{Scoop}

The momentum integral corresponding to Fig.\ \ref{fig6} can be
evaluated directly from Eq.\ (\ref{eq13}):
\begin{equation}
{\cal Q}=\int{\cal D}k\,\frac{i}{(k^2-m^2)}
=\frac{\xi^\epsilon\Gamma(-1+\epsilon)}{(4\pi)^2}m^2.
\label{eq16}
\end{equation}
${\cal Q}$ is independent of the incoming momentum, and
proportional to the mass in the loop. Hence, Goldstone boson {\em
Scoop}s are always zero ($m_w=0$) and have simply been ignored
here. For the case of a Higgs scoop with a self energy insertion,
we also need
\begin{equation}
{\cal B}_Q=i\frac{\partial{\cal Q}}{\partial M_H^2}=
\frac{i\xi^\epsilon\Gamma(\epsilon)}{(4\pi)^2}.
\label{eq17}
\end{equation}
For the $\bf w$, ${\cal Q}$
vanishes identically, and ${\cal B}_Q=0$ as well.

\subsubsection{Bubbles}

The momentum integral for the bubble diagrams is given in the
convention in Fig.\ \ref{fig6} by:
\begin{equation}
{\cal B}=\int{\cal D}k\,\frac{i}{(k^2-a^2)}\,
\frac{i}{((k+q)^2-b^2)}.
\label{eq18}
\end{equation}
We will need ${\cal B}$ to order $\epsilon$,
an order higher than retained in most calculations.
Combining the denominators using a Feynman parameter $x$ and
performing the ${\cal D}k$ integration gives
\begin{equation}
{\cal B}=\frac{-i\Gamma(\epsilon)\xi^\epsilon}{(4\pi)^2}
\int_0^1dx\,\left[\hat a^2(1-x)+\hat b^2x-\hat s
X\right]^{-\epsilon},
\label{eq19}
\end{equation}
where $X=x(1-x)$. Similarly,
\begin{equation}
\partial{\cal B}=\frac{\partial{\cal B}}{\partial q^2}
=\frac{-i\Gamma(1+\epsilon)\xi^\epsilon}{(4\pi)^2 M_H^2}
\int_0^1dx\,X\left[\hat a^2(1-x)+\hat b^2x-\hat s
X\right]^{-1-\epsilon}.
\label{eq20}
\end{equation}

The cases we will need correspond to the choices
\[
\begin{array}{ll}
{\cal B}_0: &\hat s=1,\quad \hat a=\hat b=0;\\
{\cal B}_1: &\hat a=1,\quad \hat b=\hat s=0;\\
{\cal B}_2: &\hat a=\hat b=\hat s=1.
\end{array}
\]

If $\hat a=\hat b=0$ or if $\hat s=0$, the integrals give beta
functions and the answers can be written down immediately, as can
the derivatives $\partial{\cal B}$ for the same cases.
For ${\cal B}_2$, we expand
the denominator in powers of $\epsilon$.
The result
at overall order $\epsilon$
involves the integrals:
\begin{equation}
\int dx\ln(1-x+x^2);\quad \int dx\ln^2(1-x+x^2);\quad
\int dx\frac{\ln(1-x+x^2)}{1-x+x^2}.
\label{eq21}
\end{equation}
 These  can be
evaluated using the factorization
\begin{equation}
1-x+x^2=(1-c_+x)(1-c_-x),\qquad c_\pm=e^{\pm i\pi/3}.
\label{eq22}
\end{equation}
The last two integrals then lead to dilogarithms and, in
particular, to the imaginary part of dilogarithms, Clausen
functions \cite{dilog}.
These
integrals contribute such
characteristic quantities as
${\bf C}={\rm Cl}(\pi/3)=1.01494\cdots$,
$\pi\sqrt{3}$, and $\pi\sqrt{3}\ln3$.
DeVoto and Duke \cite{devoto} give a number of results which are
useful in these and later calculations in an extensive
compilation. Further results are given in Appendix D.

We will also need
${\cal B}_0(s)=s^{-\epsilon}
{\cal B}_0$ and
${\cal B}_2(s)$ for $s=p^2 \not= M_H^2$.
The second function can again be
evaluated by expansion in powers of $\epsilon$, or by using a
method based on a Barnes'-type representation for the denominator
in Eq.\ (\ref{eq20}). We have found the second method---which is
apparently new in this context---to be a very powerful way of
handling the integrals over Feynman parameters encountered in
dimensional regularization.  The method is discussed in detail in
Appendix B.

\subsubsection{Triangle}

The {\em Triangle} graph in Fig.\ \ref{fig6} is related to the
{\em Bubble} by ${\cal T}=i\partial{\cal B}/\partial a^2$,
\begin{equation}
{\cal T}=\int{\cal D}k\left(\frac{i}{(k^2-a^2)}\right)^2\,
\frac{i}{((k+q)^2-b^2)}.
\label{eq23}
\end{equation}
Hence, from Eq.\ (\ref{eq19}),
\begin{equation}
{\cal T}=i\frac{\partial{\cal B}}{\partial a^2}
=\frac{-\Gamma(1+\epsilon)\xi^\epsilon}{(4\pi)^2 M_H^2}
\int_0^1dx(1-x)\left[\hat a^2(1-x)+\hat b^2x-\hat
sX\right]^{-1-\epsilon},
\label{eq24}
\end{equation}
and
\begin{equation}
\frac{\partial{\cal T}}{\partial q^2}=
\frac{-\Gamma(2+\epsilon)\xi^\epsilon}{(4\pi)^2 M_H^2}
\int_0^1dx\,(1-x)X\left[\hat a^2(1-x)+\hat b^2x-\hat
sX\right]^{-2-\epsilon}.\label{eq25}
\end{equation}

There are four cases of interest here:
\[
\begin{array}{ll}
{\cal T}_0: &\hat s=1,\quad \hat a=\hat b=0;\\
{\cal T}_1: &\hat b=1,\quad \hat a=\hat s=0;\\
{\cal T}_2: &\hat a=1,\quad \hat b=\hat s=0;\\
{\cal T}_3: &\hat a=\hat b=\hat s=1.
\end{array}
\]
The first three lead to beta functions, as for the analogous
bubble
${\cal
B}_i$. By power counting, or by comparison to the analogous ${\cal
B}_i$, it can be seen that the divergences of ${\cal T}_0$ and
${\cal T}_1$ are infrared divergences. ${\cal T}_3$ was evaluated
by expanding the integrand in powers of $\epsilon$ following the
method used for ${\cal B}_2$.  We will also need ${\cal T}_0(s) =
s^{-1-\epsilon} {\cal T}_0$ and ${\cal T}_3 (s)$. The latter was
evaluated both by the Barnes' method in Appendix B and by direct
expansion.

\subsection{The two-loop diagrams}
\subsubsection{Sunset}

The {\em Sunset} graph in Fig.\ \ref{fig7} is one of two new
topologies at this order, and is the only two-loop
self-energy graph in a
symmetric $\phi^4$ theory other than the double {\em Scoop}s.
The momentum integral is given in the conventions of Fig.\
\ref{fig7} by
\begin{equation}
{\cal S}=\int{\cal D}k{\cal D}\ell\,
\frac{i}{(k^2-a^2)}\,\frac{i}{((k-\ell+q)^2-b^2)}\,
\frac{i}{(\ell^2-c^2)}.
\label{eq26}
\end{equation}
Combining the first two denominators using a Feynman parameters
$x$ and performing the $k$ integration, then combining the
remaining denominators using a second parameter $y$ and
integrating over $\ell$ gives the expressions
\begin{equation}
{\cal
S}=\frac{-iM_H^2\xi^{2\epsilon}\Gamma(-1+2\epsilon)}{(4\pi)^4}
\int_0^1 dx\, X^{-\epsilon} \,
\int_0^1 dy\, y^{\epsilon-1}
\left[\hat c^2(1-y)+\hat\alpha^2y-\hat
sy(1-y)\right]^{1-2\epsilon},\label{eq27}
\end{equation}
and
\begin{equation}
\frac{\partial{\cal S}}{\partial q^2}
=\frac{-i\xi^{2\epsilon}\Gamma(2\epsilon)}{(4\pi)^4}
\int_0^1 dx\, X^{-\epsilon} \,
\int_0^1 dy\, y^\epsilon (1-y)
\left[\hat c^2(1-y)
+\hat\alpha^2y-\hat sy(1-y)\right]^{-2\epsilon},
\label{eq28}
\end{equation}
where
$\hat\alpha^2 = \alpha^2/ M_H^2$ with
\begin{equation}
\alpha^2=\frac{a^2}{x}+\frac{b^2}{1-x}.
\label{eq29}
\end{equation}

The four cases of interest here are
\[
\begin{array}{ll}
{\cal S}_0: &\hat a=\hat b=\hat c=\hat s=0;\\
{\cal S}_1: &\hat c=\hat s=1,\quad \hat a=\hat b=0;\\
{\cal S}_2: &\hat a=\hat b=1,\quad \hat c=\hat s=0;\\
{\cal S}_3: &\hat a=\hat b=\hat c=\hat s=1.
\end{array}
\]

The ${\cal S}_0$ case is curious because there is no internal scale
in the graph: all the masses are zero and the integral is to be
evaluated at $q^2=s=0$. For vanishing internal masses, ${\cal S}_0$
is proportional to $\hat s^{1-2\epsilon}$ multiplied by finite
integrals, hence vanishes for $\hat s\rightarrow0$. The case of
$\partial{\cal S}_0$ is more delicate since $\partial{\cal
S}_0\propto\hat s^{-2\epsilon}$ and there is a final infrared
divergence unless $\epsilon<0$. We will adopt this limit, with
$\epsilon\rightarrow 0^-$. Alternatively, to avoid this problem, we
note that ${\cal S}_0$ can also be rewritten in a form which
contains a mass scale and avoids conflicts between the infrared and
ultraviolet regularizations by using the identity \begin{eqnarray}
\frac{i}{k^2} &=& \frac{(k^2-m^2)}{(k^2-m^2)}\,\frac{i}{k^2}
\nonumber\\ &=&
\frac{i}{(k^2-m^2)}-\frac{m^2}{i}\,\frac{i}{(k^2-m^2)}\,
\frac{i}{k^2}. \label{eq30} \end{eqnarray} This substitution gives
\begin{equation} {\cal S}_0={\cal S}_1|_{q^2=0}-\frac{m^2}{i}
\int{\cal D}k{\cal D}\ell\frac{i}{k^2}\,
\frac{i}{(k^2-m^2)}\,\frac{i}{(k-\ell)^2}\, \frac{i}{(\ell+q)^2},
\label{eq31} \end{equation} with a similar expression for
$\partial{\cal S}_0$. In either case, the two contributions,
calculated independently, are identical, and both ${\cal S}_0$ and
$\partial{\cal S}_0$ vanish.

Since all lines in a {\em Sunset} graph are formally equivalent, a
judicious choice can be made of which to choose as massive in the
cases ${\cal S}_1$ and ${\cal S}_2$. The calculations simplify
enormously with the choices above. For both ${\cal S}_1$ and
${\cal S}_2$, the $x$ and $y$ integrations then factor in Eqs.\
(\ref{eq27}) and (\ref{eq28}) and lead to beta functions.

The all-massive case ${\cal S}_3$ is
troublesome,
because the last factor in Eq.\ (\ref{eq27}) mixes singularities
in the $x$
and $y$ integrations.
One useful approach, the ``$\partial p$" or ``partial p" method,
makes use of integration by parts in the original momentum-space
integral
in Eq.\ (\ref{eq26}) and in the modified integral over Feynman
parameters which results to make the divergences explicit.
It is then possible to expand the remaining integrand in powers
of $\epsilon$.
The calculation
proceeds along the general lines sketched for this case by Ramond
\cite{ramond}, but a complete calculation {\em including} finite
parts is what is needed here. The details are given in
\cite{maher}.
${\cal S}_3$ can also be calculated directly using the Barnes'
representation technique discussed in Appendix B after a factor
$(y/ X)^{1-2\epsilon}$ is extracted from the last factor
in Eq.\ (\ref{eq27}).
For $\partial{\cal S}_3$, the original expression in
Eq.\ (\ref{eq28}) is sufficient, though the Barnes method is also
simple.

\subsubsection{Acorn}

As is evident from Fig.\ \ref{fig7}, the {\em Acorn} diagram has the
character of a self energy bubble with a renormalized coupling. The
${\cal A}_Q$ diagram in Fig.\ \ref{fig7} is a {\em Scoop} with a
self-energy insertion. The two types of diagrams give Feynman
integrals of the same form with the restriction $q=0$ for ${\cal
A}_Q$. The common momentum integral is \begin{equation} {\cal
A}=\int{\cal D}k{\cal D}\ell\,
\frac{i}{(k^2-a^2)}\,\frac{i}{((k-\ell)^2-b^2)}\,
\frac{i}{(\ell^2-c^2)}\, \frac{i}{((\ell+q)^2-d^2)}. \label{eq32}
\end{equation} Combining the propagators of the ${\cal D}k$ loop and
integrating, \begin{equation} {\cal
A}=\frac{i^2(4\pi\mu^2)^\epsilon\Gamma(2-\omega)}{(4\pi)^2}
\int_0^1\frac{dx}{X^{2-\omega}} e^{-i\pi\omega} \int{\cal D}\ell\,
\frac{i}{(\ell^2-\alpha^2)^{2-\omega}}\, \frac{i}{(\ell^2-c^2)}\,
\frac{i}{((\ell+q)^2-d^2)} \label{eq33} \end{equation} where
$\alpha^2$ is defined in Eq.\ (\ref{eq29}). Introducing two more
Feynman parameters and performing the ${\cal D}\ell$ integration, we
obtain the expression
\begin{eqnarray}
{\cal A} &=&
\frac{-\xi^{2\epsilon}\Gamma(2\epsilon)}{(4\pi)^4}
\int_0^1\frac{dx}{X^\epsilon}\, \int_0^1 dy\,y^\epsilon\, \int_0^1
dz\, z^{\epsilon-1}\nonumber \\
&&\times \left[\hat c^2y+\hat
d^2(1-y)-\hat sy(1-y)-(\hat c^2-\hat\alpha^2)
yz\right]^{-2\epsilon}. \label{eq34}
\end{eqnarray}

The particular {\em Acorn} graphs which we need correspond to the
following sets of parameters grouped according to the diagrams
in Figs.\ 2 and 3:
\[
\begin{array}{l@{:\quad}l}
{\cal A}_4 & \hat a=\hat b=\hat c=\hat d=\hat s=1;\\
{\cal A}_{2m} & \hat c=\hat d=\hat s=1,\quad \hat a=\hat b=0;\\
{\cal A}_{1m} &\hat a=\hat s=1,\quad \hat b=\hat c=\hat d=0;\\[1ex]
{\cal A}_3 & \hat a=\hat b=\hat c=1,\quad \hat d=\hat s=0;\\
{\cal A}_{20} & \hat b=\hat d=1,\quad \hat a=\hat c=\hat s=0;\\
{\cal A}_{10} & \hat c=1,\quad \hat a=\hat b=\hat d=\hat
s=0;\\[1ex] {\cal A}_{4Q} & \hat a=\hat b=\hat c=\hat d=1,\quad
\hat s=0;\\ {\cal A}_{2Q} & \hat c=\hat d=1,\quad \hat a=\hat
b=\hat s=0;\\ {\cal A}_{1Q} & \hat a=1;\quad \hat b=\hat c=\hat
d=\hat s=0. \end{array}
\]

The integral for ${\cal A}_{10}$ factors and gives a product
of beta functions. The remaining integrals can be evaluated by
writing the last factor in Eq.\ (\ref{eq34}) as
$[\cdots]=[P+Qz]$ and
integrating by parts with respect to $z$ to
eliminate the $z^{\epsilon-1}$ singularity:
\begin{equation}
{\cal A}=\frac{-\xi^{2\epsilon}\Gamma(2\epsilon)}{(4\pi)^4}
\int_0^1\frac{dx}{X^\epsilon}\,
\int_0^1 dy\,y^\epsilon\left[
\frac{1}{\epsilon(P+Q)^{2\epsilon}}+2
\int_0^1 dz\,
\frac{z^\epsilon Q}{[P+Qz]^{1+2\epsilon}}\right].
\label{eq35}
\end{equation}
We can now expand the factor $z^\epsilon$ in the last integrand to
O($\epsilon$) and
evaluate the remaining integral on $z$, and find that
\begin{equation}
{\cal A}=\frac{-\xi^{2\epsilon}\Gamma(2\epsilon)}{(4\pi)^4}
\int_0^1\frac{dx}{X^\epsilon}\,
\int_0^1 dy\,y^\epsilon
\left[\frac{1}{\epsilon P^{2\epsilon}}+2\epsilon{Li}_2\left(
\frac{-Q}{P}\right)+{\rm O}(\epsilon^2)\right],
\label{eq36}
\end{equation}
where ${\rm Li}_2(z)$ is the dilogarithm function \cite{dilog}, and
\begin{equation}
P=\hat c^2y+\hat d^2(1-y)-\hat sy(1-y),\quad
Q=-y(\hat c^2-\hat\alpha^2).
\label{eq37}
\end{equation}
In most cases the dilogarithm needs to be integrated by parts with
respect to $y$. This creates a set of awkward but finite integrals.

Differentiating of the expression in Eq.\ (\ref{eq36}) with
respect to $s$ gives an expression for $\partial{\cal A}$ with
much the same character,
\begin{equation}
\partial{\cal A}=-\frac{\xi^{2\epsilon}\Gamma(1+2\epsilon)}%
{(4\pi)^4M_H^2}\int_0^1\frac{dx}{X^\epsilon}\,
\int_0^1 dy\,y^{1+\epsilon}
(1-y)\left[\frac{1}{\epsilon P^{1+2\epsilon}}-\frac{1}{P}
\ln\left(\frac{P+Q}{P}\right)+{\rm O}(\epsilon)\right].
\label{eq38}
\end{equation}

The ${\cal A}_Q$ graphs are {\em Scoop}s with a particle exchange
across the loop;
$\partial {\cal A}_Q$ therefore vanishes identically.
For ${\cal A}_4$, $P=1-y(1-y)$ and, as for ${\cal
B}_2$, the $1/P^{2\epsilon}$ term in Eq.\ (\ref{eq36}) must be
expanded to O($\epsilon^2$),
and leads to integrals related to dilogarithms.
The same term appears in ${\cal
A}_{2m}$. In all other cases,
the
$1/P^{2\epsilon}$ term leads to a
beta function when integrated, and the hardest part of the
calculation is the extraction of the finite pieces.

We will also need ${\cal A}_{1m}, \partial {\cal A}_{1m}/
\partial a^2$, $\partial {\cal A}_{1m}/\partial b^2,
\partial {\cal A}_{1m}/\partial d^2, {\cal A}_{2m},
\partial {\cal A}_{2m}/\partial a^2$
and $\partial {\cal A}_{2m}/\partial d^2$, all as
functions of $s$, for our calculation of the {\em Lemon} graphs
${\cal L}_1, {\cal L}_{2m}$, and their derivatives. These
integrals were all evaluated using the Barnes' representation
method.
The results of the {\em Acorn} calculations are summarized in
Appendix A.

\subsubsection{Eye-in-the-Sky}

The {\em Eye-in-the-Sky} diagrams in Fig.\ \ref{fig7} are simply
self-energy bubbles with self-energy bubble insertions. The most
singular terms from these diagrams will cancel exactly against the
mass counterterm and {\em Scoop} insertions. The generic momentum
integral is
\begin{equation}
{\cal E}=\int{\cal D}k\,{\cal D}\ell\,
\frac{i}{(k^2-a^2)}\,\frac{i}{((k-\ell)^2-b^2)}\,
\left(\frac{i}{(\ell^2-c^2)}\right)^2\,
\frac{i}{((\ell+q)^2-d^2)},
\label{eq39}
\end{equation}
and is simply a derivative of an {\em Acorn} integral,
${\cal E}=i\partial{\cal A}/\partial c^2$.
Thus, from Eq.\ (\ref{eq34}),
\begin{eqnarray}
{\cal E} &=&
\frac{i\xi^{2\epsilon}\Gamma(1+2\epsilon)}{(4\pi)^4M_H^2}
\int_0^1 dx\, X^{-\epsilon} \int_0^1 dy\, y^{1+\epsilon}
\int_0^1 dz\, z^{\epsilon-1}(1-z)\nonumber\\
&&\times\left[\, \hat c^2 y + \hat d^2 (1-y) -
\hat s y (1-y) - (\hat c^2 - \hat\alpha^2) yz\,
\right]^{-1-2\epsilon}.
\label{insert40}
\end{eqnarray}
Alternatively,
\begin{eqnarray}
{\cal E}
&=&\frac{i\xi^{2\epsilon}\Gamma(1+2\epsilon)}{(4\pi)^4M_H^2}
\int_0^1\frac{dx}{X^\epsilon}\,
\int_0^1 dy\,y^{1+\epsilon}\nonumber \\
&&\times \left[\frac{1}{\epsilon P^{1+2\epsilon}}-
\left(\frac{1}{P}+\frac{1}{Q}\right)\ln\left(\frac{P+Q}{P}\right)
+{\rm O}(\epsilon)
\right],\label{eq40} \\
\partial{\cal E} &=&
\frac{i\xi^{2\epsilon}\Gamma(2+2\epsilon)}{(4\pi)^4M_H^4}
\int_0^1\frac{dx}{X^\epsilon}\,
\int_0^1 dy\,y^{2+\epsilon}(1-y)\nonumber \\
&&\times \left[
\frac{1-\epsilon}{\epsilon P^{2+2\epsilon}}-
\frac{1}{P^2}\ln\left(\frac{P+Q}{P}\right)+{\rm O}(\epsilon )
\right],
\label{eq41}
\end{eqnarray}
when these expressions are well-defined.

We will need the same six sets of parameters as appeared for the
ordinary {\em Acorn} diagrams:
\[
\begin{array}{l@{:\quad}l}
{\cal E}_5=\partial_{c^2}{\cal A}_4 & \hat a=\hat b=\hat c=
\hat d=\hat s=1;\\
{\cal E}_3=\partial_{c^2}{\cal A}_{2m} &
\hat c=\hat d=\hat s=1,\quad \hat a=\hat b=0;\\
{\cal E}_1=\partial_{c^2}{\cal A}_{1m} &
\hat a=\hat s=1,\quad \hat b=\hat c=\hat d=0;\\[1ex]
{\cal E}_4=\partial_{c^2}{\cal A}_3 &
\hat a=\hat b=\hat c=1,\quad \hat d=\hat s=0;\\
{\cal E}_2^*=\partial_{c^2}{\cal A}_{10} &
\hat c=1,\quad \hat a=\hat b=\hat d=\hat s=0;\\
{\cal E}_2=\partial_{c^2}{\cal A}_{20} &
\hat b=\hat d=1,\quad \hat a=\hat c=\hat s=0.
\end{array}
\]
These graphs display characteristics similar to the parent {\em
Acorn} graphs. Thus,
${\cal E}_2^*=i\partial{\cal A}_{10}/\partial c^2$ was
calculated using  the original Feynman parametrization
in Eq.\ (\ref{insert40}).
${\cal E}_5$ and ${\cal E}_3$ were calculated
by expanding
the factor $P^{-2\epsilon}$
terms  in powers of $\epsilon$.
No expansion was necessary for ${\cal E}_2^*$ or
${\cal E}_4$.
${\cal E}_1$ was  calculated as a function of $\hat s$
and for $\hat s=0$
using the Barnes' technique.

${\cal E}_2=i\partial{\cal A}_{20}$ was calculated by a different
route. The problem here is that the $\ln P/P$ term is
$\ln(1-y)/(1-y)$, i.e., divergent. The
original
momentum integral was
rewritten using
partial fractions, as
\begin{equation}
\frac{i}{((\ell+q)^2-M_H^2)}\,\frac{i}{\ell^2}
=\frac{i}{M_H^2}\left(
\frac{i}{(\ell+q)^2-M_H^2}-\frac{i}{\ell^2} \right)
+\frac{i^2}{M_H^2}\,\frac{2\ell\cdot
q+q^2}{\left( (\ell+q)^2-M_H^2 \right)\ell^2}.
\label{eq43}
\end{equation}
The last term is proportional to $q^2$ after integration, so
vanishes in the limit $q^2=0$ needed for ${\cal E}_2$. The first
two terms give {\em Acorn\/} integrals, and ${\cal E}_2$ therefore
splits
into a difference of known terms,
\begin{equation}
{\cal E}_2=\frac{i}{M_H^2}({\cal A}_{20}-{\cal A}_{1Q}).
\label{eq44}
\end{equation}
$\partial {\cal E}_2$ was calculated using the Barnes method.

\subsubsection{Lemon}

The {\em Lemon} topology shown in Fig.\ \ref{fig7} was by far the
most troublesome, and three very different techniques were used to
handle these diagrams
\cite{kreimer}. In addition, four separate numbers were
ultimately extracted by
numerical evaluation of two-dimensional
integrals  using the program {\tt VEGAS}
\cite{vegas}. The momentum integral is
\begin{equation}
{\cal L}=\int{\cal D}k\,{\cal D}\ell
\frac{i}{(k^2-a^2)}\,
\frac{i}{((k+q)^2-b^2)}\,
\frac{i}{((k-\ell)^2-c^2)}\,
\frac{i}{(\ell^2-d^2)}\,
\frac{i}{((\ell+q)^2-e^2)}.
\label{eq46}
\end{equation}
This integral is ultraviolet convergent, but one case,
$\partial{\cal L}_{20}$ is infrared divergent. The parameter sets
needed are:
\[
\begin{array}{l@{:\quad}l}
{\cal L}_5 & a^2=b^2=c^2=d^2=e^2=s=M_H^2;\\
{\cal L}_{2m} & a^2=b^2=s=M_H^2,\quad c^2=d^2
=e^2=0;\\
{\cal L}_1 & c^2=s=M_H^2,\quad a^2=b^2=d^2
=e^2=0;\\[1ex]
{\cal L}_{20} & a^2=e^2=M_H^2,\quad b^2=c^2
=d^2=s=0.\\
{\cal L}_3 & a^2=c^2=d^2=M_H^2,\quad
b^2=e^2=s=0;
\end{array}
\]

The momentum integral can be performed formally using Feynman
parameters.
Combining the $a$ and $b$ propagators first, adding in
the $c$ propagator and integrating over $k$, then combining
the result with the remaining $d$ and $e$ propagators in
one step and integrating over $\ell$ gives
\begin{equation}
{\cal L} = \frac{i\xi^{2\epsilon}\Gamma(1+2\epsilon)}%
{(4\pi)^4 M_H^2} \int_0^1 dx\, x X^\epsilon\,
\int_0^1
 dy\, \int_0^1 du\, dv\, dw\,
u^\epsilon\, \delta(1-u-v-w)\left( M - \hat s S \right)^{-1-
2\epsilon}
\label{insert48}
\end{equation}
and
\begin{equation}
\partial{\cal L} = \frac{i\xi^{2\epsilon}\Gamma(2+2\epsilon)}%
{(4\pi M_H)^4} \int_0^1 dx\, xX^\epsilon\,
\int_0^1 dy\,\int_0^1 du\, dv\, dw\,
u^\epsilon \delta(1 - u - v - w)
\frac{S}{\left( M - \hat s S\right)^{2+2\epsilon}},
\label{insert49}
\end{equation}
where
\begin{eqnarray}
M &=& u\left[ \hat a x(1 - y) + \hat b xy + \hat c
(1 - x)\right] + \hat d vX + \hat e wX,\nonumber\\
S &=& ux^2y(1 - y) + X(uy + w)(1 - uy - w).
\label{insert50}
\end{eqnarray}

The Feynman parameter integrals are generally too
complicated for direct integration except for the $\Pi_w$ cases
where $s=0$ and just mass terms survive.
In particular, ${\cal L}_{20}$,
$\partial{\cal L}_{20}$ and ${\cal L}_3$ can all be calculated
analytically. $\partial{\cal L}_3$ was calculated partially
analytically and partially by a two-dimensional numerical
integration. However, the Feynman parameter expressions are not
generally well-suited to numerical integration. For example, the
expression for ${\cal L}_3$ had to be reduced analytically to a
two dimensional integral before the
result obtained by
numerical integration agreed with
the analytic integration. This is presumably because {\tt VEGAS}
could not adapt to the complicated singularity structure.

A more useful technique for ${\cal L}_5$, the all-massive case,
was based on the hyperspherical formalism \cite{roskies}. In this
technique, a Euclidean propagator is expanded in Gegenbauer
polynomials $C_n^1$ (or equivalently, Chebyshev polynomials of the
first kind $U_n$):
\begin{equation}
\frac{1}{(k-\ell)^2+m^2}=
\frac{Z_{KL}}{KL}\sum_{n=0}^\infty Z_{KL}^n
C^1_n(\hat K\cdot\hat L),
\label{eq47}
\end{equation}
where $K$, $L$ and $\hat K$, $\hat L$ are the magnitudes and
four-dimensional unit vectors of
the Euclidean momenta $k$, $\ell$  and
\begin{eqnarray}
Z_{KL}&=&\frac{K^2+L^2+m^2-\sqrt{(K^2+L^2+m^2)^2-4K^2L^2}}{2KL},
\quad m\not=0,
\nonumber\\
&=& {\rm min}\,\left(\frac{K}{L},\frac{L}{K}\right)\quad
\mbox{for } m^2=0.
\label{eq48}
\end{eqnarray}
The angular integrals are easily carried out for planar graphs
using the orthogonality relations
\begin{equation}
\int C_n^1(\hat a\cdot\hat b)C_m^1(\hat b\cdot\hat c)\,
\frac{d\Omega_b}{2\pi^2}=\frac{\delta_{mn}}{n+1}\,
C^1_n(\hat a\cdot\hat c).
\label{eq49}
\end{equation}

Only propagators with angular dependence need be expanded, i.e.,\
for the {\em Lemon}, the propagators of the $b$, $c$ and
$e$ lines (see Fig.\ \ref{fig7}). The result for ${\cal L}_5$
after
performing
the angular integrations, resumming
the series, and making the analytic continuation back to Minkowski
space, is
\begin{equation}
{\cal L}_5=\frac{-i}{Q^2}\left.\int_0^\infty
\frac{dK^2}{(4\pi)^2}\,
\int_0^\infty \frac{dL^2}{(4\pi)^2}
\frac{\ln(1-Z_{KQ} Z_{KL} Z_{LQ})}{(K^2+a^2)(L^2+
d^2)}\right|_{Q^2=-M_H^2}.
\label{eq50}
\end{equation}
There is no known change of variables that will reduce this
case to an analytically integrable form \cite{roskies}.
Equation (\ref{eq50}) was integrated numerically instead. The
advantage of the hyperspherical formalism
is that the integral has been reduced from four to
two dimensions, and the integrand is a much smoother function.

The remaining integrals ${\cal L}_1,\ \partial {\cal L}_1,\ {\cal
L}_{2m},\ \partial {\cal L}_{2m}$, and $\partial {\cal L}_5$ are all
finite, but numerical integration of either the Feynman parameter or
hyperspherical expressions is complicated by singularities or, in
the hyperspherical case, a complicated complex analytic structure.
For these a third technique was used, Kotikov's method of
differential equations \cite{kotikov}. Kotikov uses the method to
find a first order differential equation in the particle mass
with, hopefully, ``easily calculable" inhomogeneous terms. In the
present case, we obtain  a differential equation in $\hat s$
(the only free variable is the ratio $\hat s=s/M_H^2$) by taking
the derivative of the momentum integral in Eq.\ (46) with respect
to the incoming momentum. The equation can be simplified by using
the six identities \cite{thooft}
%
% equation
%
\begin{equation}
\label{insert54}
	\int dk\int d\ell\frac{\partial }{\partial h_\mu }\left[ p_\mu
	(\cdots) \right]=0
\end{equation}
where $(\cdots)$ is a product of propagators, and $p_\mu $ is any
of the variables $k_\mu ,\ell_\mu $, and $q_\mu $, and $h_\mu $ is
either $k_\mu $ or $\ell_\mu $.

The result is an inhomogeneous first order differential equation
with the inhomogeneous term
expressible in terms of {\em Bubbles\/}, {\em
Triangles\/}, {\em Acorns\/}, and their derivatives. ${\cal L}$
can then be found by direct integration. This is an interesting
technique, and the remaining cases were handled this way as
discussed in Appendix C. Our exposition there is less general than
Kotikov's \cite{kotikov} since we treat only specific diagrams.

\subsection{The self-energy terms and renormalization
constants}\label{self-energy}

We will write the expansions for the self-energy terms and wave
function renormalization constants in powers of the coupling
$\lambda$ as
\begin{eqnarray}
\Pi^0 &=& \Pi^{(1)}+\Pi^{(2)}+\cdots\nonumber\\
Z &=& 1+Z^{(1)}+Z^{(2)}+\cdots
\label{eq52}
\end{eqnarray}
The one-loop self energies are given   by the first
two terms in Eqs.\ (\ref{eq11}) and (\ref{eq12}) with the wave
function and coupling renormalizations ignored,
\begin{equation}
\Pi_w^{(1)}(0)=-\frac{3\lambda}{16\pi^2}M_H^2\xi^\epsilon
\left(\,
\frac{1}{\epsilon}+(1-\gamma)+\frac{\epsilon }{2}(2-2\gamma+\gamma^2
+\zeta(2))\,
+{\rm O}(\epsilon^2)
\right)
\label{eq53}
\end{equation}
and
\begin{eqnarray}
\Pi_H^{(1)}(M_H^2) &=& -\frac{3\lambda}{16\pi^2}M_H^2\xi^\epsilon
\left[\,
\frac{5}{\epsilon}+(9-\pi\sqrt{3}-5\gamma)\right.\nonumber\\
&& +\epsilon \left(\frac{5}{2}\gamma^2-9\gamma+17-\frac{\pi^2}{4}-
2\pi\sqrt{3}(1-\case{1}/{2}\gamma)+
\pi\sqrt{3}\ln 3-4\sqrt{3}{\bf C}\right)\nonumber\\
&&\left. + i\pi +(2-\gamma)i\pi\epsilon\phantom{\frac{1}{2}}
+{\rm O}(\epsilon^2)
\right].
\label{eq54}
\end{eqnarray}
Here $\gamma=0.5772\cdots$ is Eulers constant, $\zeta(2)=\pi^2/6$,
and ${\bf C}=C\ell\left(\frac{\pi}{3}\right)=1.01494\cdots$.
We have not expanded the factors $\xi^\epsilon$ for simplicity.
This will also make the disappearance of the arbitrary scale $\mu$
manifest in our final results: the complete cancellation of the
poles in $\epsilon$ will allow us to take the limit
$\epsilon\rightarrow0$,
$\xi^\epsilon=(4\pi\mu^2/M_H^2)^\epsilon
\rightarrow1$ with no pieces left
over. The one-loop contributions to the wave function
renormalization constants are similarly given by
\begin{equation}
Z_w^{(1)}=-\frac{\lambda}{16\pi^2}\xi^\epsilon\left[1+
\epsilon (\case{3}/{2}-\gamma)+{\rm O}(\epsilon ^2)\right],
\label{eq55}
\end{equation}
and
\begin{equation}
Z_H^{(1)}=\frac{\lambda}{16\pi^2}\xi^\epsilon\left[12
-2\pi\sqrt{3}+\epsilon (24-12\gamma-3\pi\sqrt{3}(1-\case{2}/{3}\gamma)
+2\pi\sqrt{3}\ln3
-8\sqrt{3}{\bf C})+{\rm O}(\epsilon^2)\right].
\label{eq56}
\end{equation}
The leading terms in these expressions agree with the results of
previous one-loop calculations
\cite{djl1,djl2,dw1,pass1,dw2,pass2,mw}.

The real parts of the self energies at O($\lambda^2$) involve the
corrections to the one-loop self energies from the one-loop wave
function and coupling renormalizations as well as the new two-loop
contributions. The algebraic programming system {\tt REDUCE}
was used to manipulate the expressions in Eqs.\
(\ref{eq11}) and (\ref{eq12}) and obtain the expansions needed.
The results, evaluated on mass shell, are
\begin{eqnarray}
\Pi_w^{(2)}(0) &=
{\displaystyle\frac{\lambda^2}{(16\pi^2)^2}}\xi^{2\epsilon}
M_H^2&\Biggl(\frac{-45}{\epsilon^2}+\frac{6}{\epsilon}
(-22+15\gamma+3\sqrt{3}\pi)\nonumber\\
&&-72\sqrt{5}\ln\left(\frac{\sqrt{5}+1}{2}\right)-24\ln^2
\left(\frac{\sqrt{5}+1}{2}\right)+32\ln2-90\gamma^2
\nonumber\\
&& +264\gamma+40\zeta(2)-\frac{753}{2}+66\sqrt{3}{\bf C}
-36\sqrt{3}\pi\gamma+45\sqrt{3}\pi\nonumber\\
&&-18\pi\sqrt{3}\ln 3+{\rm O}(\epsilon)
\Biggr).
\label{eq57}
\end{eqnarray}
and
\begin{eqnarray}
{\rm Re}\,\Pi_H^{(2)}(M_H^2) & = &
{\displaystyle\frac{\lambda^2}{(16\pi^2)^2}}
\xi^{2\epsilon}M_H^2
\biggl(\frac{-189}{\epsilon^2}+\frac{9}{\epsilon}
(-71+42\gamma+10\sqrt{3}\pi)\nonumber\\
&& -144\ln2-45\pi^2-378\gamma^2
\nonumber\\
&& -36 K_2 -162 K_5+1278\gamma
-63\zeta(2)+90\zeta(3)\nonumber\\
&&-1524+126\sqrt{3}{\bf C}-180\sqrt{3}\pi\gamma+408\sqrt{3}\pi
\nonumber\\
&&-81\xi(2)\ln2-90\pi\sqrt{3}\ln 3+{\rm O}(\epsilon)
\biggr).
\label{eq58}
\end{eqnarray}
The constants $K_i$ are from numerical integrations, and are given
in Appendix \mbox{A}. These numbers were calculated to accuracies
which vary from a part in $10^4$ to a part in $10^6$,
more than sufficient for our
purposes despite the large topological weights and the
cancellations which occur.

The O($\lambda^2$) contributions to the wave function
renormalization constants obtained from the derivatives of Eqs.\
(\ref{eq11}) and (\ref{eq12}) using {\tt REDUCE} are
\begin{eqnarray}
Z_w^{(2)} & =
{\displaystyle\frac{\lambda^2}{(16\pi^2)^2}}\xi^{2\epsilon} &
\Biggl(-\frac{3}{\epsilon}+42\sqrt{5}\ln
\left(\frac{\sqrt{5}+1}{2}\right)-96\ln^2\left(
\frac{\sqrt{5}+1}{2}\right)\nonumber\\
&& +\frac{400}{9}\ln2+6\gamma-\frac{38}{3}\zeta(2)
-12K_3-\frac{1525}{54}+3\sqrt{3}\pi+{\rm O}(\epsilon)\Biggr)
\label{eq59}
\end{eqnarray}
and
\begin{eqnarray}
Z_H^{(2)} & = &
{\displaystyle\frac{\lambda^2}{(16\pi^2)^2}}\xi^{2\epsilon}
\left(-\frac{3}{\epsilon}+144\ln 2 +42 \pi^2 +
 81\zeta(2) \ln 2\right.\nonumber\\
&&+\,36K_2 + 162 K_5
+6\gamma +33\zeta(2) -90\zeta(3)\nonumber\\
&& \left.-16\sqrt{3}\pi\ln 3 + 334\sqrt{3}{\bf C} +\frac{273}{2}
 -292\sqrt{3}\pi
+{\rm O}(\epsilon)
\right).
\label{eq60}
\end{eqnarray}
The leading $1/\epsilon$ divergences in $Z_w^{(2)}$ and $Z_H^{(2)}$
are identical because of the original SO(4) symmetry of the theory
which is preserved in the interactions at large momenta.

Finally, the bare coupling $\lambda_0$ can be identified as the
coefficient of $-\frac{1}{4}({\bf w}_0^2+H_0^2)^2$ in Eq.\
(\ref{eq7}),
\begin{equation}
\lambda_0=\frac{\lambda}{Z_w}\left(1-
\frac{{\rm Re}\,\Pi^0_H(M_H^2)-\Pi^0_w(0)}{M_H^2}\right).
\label{eq61}
\end{equation}
Substitution of the results above using {\tt REDUCE} gives
\begin{eqnarray}
\lambda_0 &=& \lambda+\frac{\lambda^2\xi^\epsilon}{16\pi^2}
\left(\frac{12}{\epsilon}+25-12\gamma-3\sqrt{3}\pi
+\epsilon\,[\,3\pi\sqrt{3}\ln 3 - 12\sqrt{3}{\bf C}
- 6\pi\sqrt{3}(1-\case{1}/{2}\gamma)\right.\nonumber\\
&& - \left.\pi^2 + 6\gamma^2 - 25\gamma +\frac{99}{2}\,]+{\rm O}
(\epsilon^2)
\right)+
\frac{\lambda^3\xi^{2\epsilon}}{(16\pi^2)^2}
\Biggl(\frac{144}{\epsilon^2}+\frac{18}{\epsilon}(29-16\gamma
-4\sqrt{3}\pi)\nonumber\\
&& + \frac{32906}{27}+12K_3+162K_5+36K_2-90\zeta(3)\nonumber\\
&& +\frac{185}{3}\zeta(2)-1044\gamma+288\gamma^2
+54\pi^2+\frac{1184}{9}\ln 2\nonumber\\
&&  + 72\ln^2\left(
\frac{\sqrt{5}+1}{2}\right)-114\sqrt{5}\ln\left(
\frac{\sqrt{5}+1}{2}\right)\nonumber\\
&& +81 \zeta (2) \ln 2
-369\pi\sqrt{3} + 144\pi\sqrt{3}\gamma -60\sqrt{3}{\bf C}
\nonumber\\
&&+72\pi\sqrt{3}\ln 3 + {\rm O}(\epsilon)
\Biggr) + {\rm O}(\lambda^4).
\end{eqnarray}

\section{THE SCATTERING AMPLITUDES}

The two-particle scattering diagrams generated by the quartic
interactions in Eq.\ (7) are shown in Fig.\ 4.  There is just one
generic one-loop diagram with an associated amplitude ${\cal
B}_S$, and two two-loop diagrams with amplitudes ${\cal D}_S$ and
${\cal A}_S$, the last named after the similar {\em Acorn}
self-energy diagram.  The small number of diagrams is the result
of the suppression of the dimension-three trilinear couplings in
the limit $\sqrt{s}\gg M_H^2$, as was noted earlier.  The
functions ${\cal B}_S$, ${\cal D}_S$, and ${\cal A}_S$ are given
in Appendix A.

We note  that $M_H$ can be
neglected on internal lines
in the high energy limit, and
that the scattering amplitudes
consequently have
the SO(4) symmetry of the bare quartic couplings.  This symmetry
is reflected in the structure of the $4\times4$ matrix
\mbox{\boldmath${\cal F}$}
of unrenormalized transition amplitudes.  All of these amplitudes
can be expressed in terms of a single function $A(s,t,u)$, where
\begin{eqnarray}
A(s,t,u) &=& -2\lambda_0-i(-i\lambda_0)^2 A^{(1)} -
i(-i\lambda_0)^3 A^{(2)},\\
A^{(1)}(s,t,u) &=& 16{\cal B}_S(s) + 4{\cal B}_S(t) + 4{\cal
B}_S(u),\\
A^{(2)}(s,t,u) &=& 104{\cal D}_S(s) + 8{\cal D}_S(t) + 8{\cal D}_S
(u)\nonumber \\
&& + 176 {\cal A}_S(s) + 80{\cal A}_S(t) + 80{\cal A}_S(u),
\end{eqnarray}
and $\lambda_0$ is the bare coupling defined in Eqs.\ (63) and
(64).
We will write
the two-body channels in the order $w^+ w^-,\ zz,\
HH,$ and
$zH$.
\mbox{\boldmath${\cal F}$} then has the structure
\begin{equation}
\mbox{\boldmath${\cal F}$} =
\left(
\begin{array}{cccc}
A(s)+A(t) & A(s) & A(s) & 0\\
A(s) & A(s) + A(t) + A(u) & A(s) & 0\\
A(s) & A(s) & A(s) + A(t)+A(u) & 0\\
0 & 0 & 0 & A(t)
\end{array} \right),
\end{equation}
where we have indicated only the first variable in $A$ since this
function is unchanged by an interchange of the remaining two
variables.

As noted after Eq.\ (62), the leading terms in
$1/\epsilon$ in the
renormalization constants $Z_W$ and $Z_H$  are
identical because of the SO(4) symmetry of the initial theory.  We
will therefore make an intermediate renormalization which
preserves the SO(4) structure by multiplying $A$ by the field
renormalization factor $(Z_W^{1/2})^4$ and replacing $\lambda_0$
in Eq.\ (65) by the expression in Eq.\ (63).  This gives
\begin{eqnarray}
A_R(s,t,u) &=& -2\lambda\left(1-
\frac{{\rm Re}\Pi^0_H(M_H^2)-\Pi^0_w(0)}{M_H^2}\right)Z_w\nonumber\\
&& +i\lambda^2 A^{(1)}
\left(1-\frac{{\rm Re}\Pi_H^{(1)}(M_H^2)-\Pi_w^{(1)}(0)}%
{M_H^2}\right)^2 + \lambda^3 A^{(2)},
\end{eqnarray}
where each term is to be expanded to O($\lambda^3$). $A_R$ is just
the physical amplitude needed to describe the scattering in the
$w^+w^-$, $zz$ sector. Channels involving external Higgs bosons
are multiplied by a finite renormalization factor $\left( Z_H/Z_w
\right)^{1/2}$ for each
external Higgs boson.

After a number of nontrivial cancellations, $A_R$ reduces in the
limit $\epsilon\rightarrow0$ to a finite expression independent of
the arbitrary scale $\mu$ which appeared at intermediate steps of
the dimensional regularization,
\begin{eqnarray}
A_R(s,t,u) &=& -2\lambda + \frac{\lambda^2}{16\pi^2}
\left(-16\ln
(-\hat s)-4\ln
(-\hat t)-4\ln
(-\hat u) + 2 +6\sqrt{3}\pi
\right)\nonumber\\
&&+\frac{\lambda^3}{(16\pi^2)^2}
\left(\begin{array}{c}
-192\ln^2(-\hat s)+176\ln(-\hat s)+96\sqrt{3}\pi\ln(-\hat s)\\
-48\ln^2(-\hat t)+80\ln(-\hat t)+24\sqrt{3}\pi\ln(-\hat t)
\\
-48\ln^2(-\hat u)+80\ln(-\hat u)+24\sqrt{3}\pi\ln(-\hat u)
\\
+60\sqrt{5}\ln\frac{\sqrt{5}+1}{2}-456\sqrt{3}%
{\bf C}+138\sqrt{3}\pi\\
+240\ln^2\frac{\sqrt{5}+1}{2}-\frac{3968}{9}\ln2
-180\pi^2\\
-72K_2
-\frac{794}{3}\zeta(2)+180\zeta(3)\\
-324K_5+24K_3+\frac{3388}{27}
-162\zeta(2)\ln2
\end{array}\right)\quad
\end{eqnarray}
where $-\hat s=e^{-i\pi}\hat s$, and
the quantities
$-\hat t$ and $-\hat u$ are real and
positive in the physical region.
Substitution of $A_R$ for $A$ in Eq.\ (69) gives the partially
renormalized transition matrix $\mbox{\boldmath${\cal F}$}_R$; the
fully renormalized matrix of (Feynman) transition amplitudes is
\begin{equation}
\mbox{\boldmath${\cal M}$}={\bf Z}\mbox{\boldmath${\cal F}$}_R
{\bf Z}
\end{equation}
where $\bf Z$ is a finite diagonal matrix of ratios of
renormalization constants,
\begin{equation}
{\bf Z}={\rm diag}(1,1,Z_H/Z_w,Z_H^{1/2}/Z_w^{1/2}),
\end{equation}
and the products in \mbox{\boldmath${\cal M}$} are to be expanded
to O($\lambda^3$).  The SO(4) symmetry of
$\mbox{\boldmath${\cal F}$}_R$ is lost in
\mbox{\boldmath${\cal M}$}.

\section{COMMENTS}

The principal results of this paper are the expressions we have
obtained for the renormalization constants $Z_w$ and $Z_H$, the
self-energy functions $\Pi _w^0(0)$ and ${\rm Re}\Pi _H^0(M_H^2)$,
the bare coupling $\lambda _0$, and the renormalized two-body
scattering amplitude $A_R(s,t,u)$ at two loops. We will use these
results in a following paper to study limits on the range of
validity of low-order perturbation theory in the Higgs sector of
the standard model.

More generally, the quantities connected with the on-mass-shell
renormalization of the theory can be used in other calculations of
$W_L^\pm,Z_L$ scattering based on the use of the equivalence
theorem, for example, in possible calculations to extend the
analysis of the scattering in the ``low-energy" limit
$M_w\ll\sqrt{s}\ll M_H$ \cite{lowenergy} to two loops. This
extension would be of considerable interest with respect to
studies of the energy at which the effects of a very massive Higgs
boson would become evident in future experiments \cite{lowenergy}.

We note finally that the new Barnes'-type method we have
introduced for factoring and evaluating integrals over Feynman
parameters has a wide range of applicability as illustrated in the
calculations above and in Appendix B. The results collected in
Appendix A and Appendix D should also be useful in future
calculations.

%%%%%%%%%%%%%%%%%%%%%%%%%%%%%%%%%%

\acknowledgements

This work was supported in part by the U.S. Department of Energy
under Contract No.\ AC02-76ER00881. One of the authors (L.D.) would
like to thank the Aspen Center for Physics for its hospitality while
parts of this work were done. We would also like to thank R.
Scharf for correspondence concerning the calculation of some of
the two-loop self-energy graphs.

\appendix
\section{RESULTS FOR THE SELF-ENERGY GRAPHS}\label{appa}

This appendix is a listing of the results of the Feynman graph
calculations. Some intermediate integrals used in the differential
equation method are also listed. Since only the real parts of the
self-energy graphs were needed for the counterterms, strict
attention was only paid to the real parts of those integrals when
expansions were made.

A factor of
$\xi^\epsilon/16\pi^2$ should be included for each loop in a
graph. Factors of $M_H^2$ should be included so that, with the
couplings given in Eqs.\ (\ref{eq11}) and (\ref{eq12}), the
self-energy graphs have dimensions of mass squared. That is,
${\cal Q},\ {\cal S}\propto M_H^2$; ${\cal B},\ {\cal A}
\propto 1$; and ${\cal T},\ {\cal E},\ {\cal L}\propto
M_H^{-2}$. The derivatives
of these functions
will have one power of $M_H^2$ less.
Note that the graphs with a $Q$ subscript have no dependence on
the external momentum, so $\partial(\mbox{graph}_Q)\equiv 0$.

Expressions which involve explicit expansions in powers of
$\epsilon$ have
been evaluated using expansions in $\epsilon$ in intermediate
steps.  These expressions are not correct beyond the highest power
of $\epsilon$ indicated.
Terms involving gamma or beta
functions have been kept in the forms given for compactness, and
are to be expanded in powers of $\epsilon$ when used.  These terms
can frequently be rearranged, and may be replaced in some cases by
quite different expressions which are equivalent to the relevant
order in $\epsilon$ if an intermediate expansion is done
differently; only the final expanded expression is meaningful.

In
the following, $B(a,b)$ is the beta function,
\begin{equation}
B(a,b)=\frac{\Gamma(a)\Gamma(b)}{\Gamma(a+b)}\ ,
\label{a1}
\end{equation}
${\rm Li}_n(x)$ is the polylogarithm \cite{dilog},
\begin{eqnarray}
{\rm Li}_n(x) &=& \int_0^x{\rm Li}_{n-1}(t)\frac{dt}{t},
\label{inserta2}\\
{\rm Li}_2(x) &=& -\int_0^x\ln(1-t)\frac{dt}{t}
=\sum_{n=1}^\infty\frac{x^n}{n^2},
\label{inserta3}
\end{eqnarray}
and
$\zeta(n)$ is the Riemann zeta function.

\begin{flushleft}
{\em Constants:}
\begin{eqnarray*}
\gamma &=& 0.57721566\ldots \\
{\bf C} &=& C\ell(\pi/3)=1.014942\cdots\nonumber\\
K_2 &=& -0.86518\\
K'_2 &=& 0.548311\\
K_3 &=& -0.1066639\\
K_5 &=& 0.9236306
\end{eqnarray*}
{\em Scattering graphs:}
\begin{eqnarray}
{\cal B}_S &=& -i (-\hat s)^{-\epsilon}\Gamma(\epsilon)
B(1-\epsilon,1-\epsilon),
\label{a2}\\
{\cal D}_S &=& ({\cal B}_S)^2,
\label{a3}\\
{\cal A}_S &=& - (-\hat s)^{-2\epsilon}
\Gamma(2\epsilon)B(2-4\epsilon,\epsilon)B(1-2\epsilon,1-2\epsilon)
B(1-\epsilon,1-\epsilon).
\label{a4}
\end{eqnarray}
{\em Scoops:}
\begin{eqnarray}
{\cal Q} &=& -\frac{\Gamma(\epsilon)}{1-\epsilon},
\label{a5}\\
{\cal B}_Q &=& -i\Gamma(\epsilon).
\label{a6}
\end{eqnarray}
{\em Bubbles:}
\begin{eqnarray}
{\cal B}_0(\hat s) &=& -i (-\hat s)^{-\epsilon}
\,\Gamma(\epsilon)\,
B(1-\epsilon,1-\epsilon),
\label{a7}\\
{\cal B}_0 &=& -ie^{i\pi\epsilon} \Gamma(\epsilon) B(1-\epsilon,
1-\epsilon)  \\
\partial {\cal B}_0 &=& -\epsilon {\cal B}_0
\label{a8}\\
{\cal B}_1 (\hat s) &=& -i\Gamma (\epsilon)
\left[1 + \epsilon \left( 2 + \frac{1}{\hat s} (1 - \hat s)
\ln (1-\hat s) \right) + {\rm O} (\epsilon^2) \right] \\
\partial {\cal B}_1 (s) &=& -i \Gamma (1 + \epsilon)
\left( \frac{1}{\hat s^2} \ln (1 - \hat s) - \frac{1}{\hat s}
+ {\rm O} (\epsilon) \right)  \\
{\cal B}_1 &=& -i\frac{\Gamma(\epsilon)}{1-\epsilon},
\label{a9}\\
\partial{\cal B}_1 &=& -i\frac{\Gamma(1+\epsilon)}{(1-\epsilon)
(2-\epsilon)},
\label{a10}\\
{\cal B}_2 (\hat s) &=& -i \Gamma (\epsilon)
\biggl( 1 - 2\epsilon \frac{4 - \hat s}{\sqrt{4 \hat s - \hat s^2}}
\arctan \frac{\hat s}{\sqrt{4 \hat s - \hat s^2}} + 2 \epsilon
 \nonumber\\
&&  + \frac{1}{2} \epsilon^2 \int_0^1 dx\, \ln^2
\left[ 1 - \hat s x (1 - x) \right] + {\rm O} (\epsilon^3) \biggr),
\\
{\cal B}_2 &=& -i\Gamma(\epsilon)\!\!\left[
1+\epsilon\left(2-\frac{\pi}{\sqrt{3}}\right)+
\epsilon^2\left(4-\frac{2\pi}{\sqrt{3}}+\frac{\pi\ln 3}{\sqrt{3}}
-\frac{4{\bf C}}{\sqrt{3}}\right)+{\rm O}(\epsilon^3)\right]\!,
\label{a11}\\
\partial{\cal B}_2 &=& i\Gamma(1+\epsilon)
\left[1-\frac{2\pi}{3\sqrt{3}}+\epsilon
\left(2-\frac{\pi}{\sqrt{3}}+\frac{2\pi\ln 3}{3\sqrt{3}}-
\frac{8{\bf C}}{3\sqrt{3}}\right)+{\rm O}(\epsilon^2)\right].
\label{a12}
\end{eqnarray}
{\em Triangles:}
\begin{eqnarray}
{\cal T}_0 (\hat s) &=& -(-\hat s)^{-1-\epsilon}
\Gamma (1+\epsilon)B(1-\epsilon,-\epsilon),\\
{\cal T}_0 &=& e^{i\pi\epsilon} \Gamma (1 + \epsilon) B
(1 - \epsilon, -\epsilon)
\label{a13}\\
\partial {\cal T}_0  &=& -e^{i\pi \epsilon }\Gamma (2+\epsilon
)B(-\epsilon ,1-\epsilon ),\\
{\cal T}_1 &=& \frac{\Gamma(\epsilon)}{1-\epsilon},
\label{a14}\\
\partial{\cal T}_1 &=& -\Gamma(2+\epsilon)B(3,-\epsilon),
\label{a15}\\
{\cal T}_2 &=& -\frac{\Gamma(1+\epsilon)}{1-\epsilon},
\label{a16}\\
\partial{\cal T}_2 &=& -\frac{\Gamma(2+\epsilon)}{(1-\epsilon)
(2-\epsilon )},
\label{a17}\\
{\cal T}_3 (\hat s) &=& - \Gamma (1 + \epsilon)
\left( \frac{2}{\sqrt{4\hat s - \hat s^2}} \arctan
\frac{\hat s}{\sqrt{4\hat s - \hat s^2}} \right. \nonumber \\
&& \left. - \epsilon \int_0^1 dx\,
\frac{x \ln[\,1 - \hat s x(1-x)\,]}{1 - \hat s x(1 - x)}
+ {\rm O} (\epsilon^2) \right), \\
{\cal T}_3 &=& -\frac{\Gamma(1+\epsilon)}{3\sqrt{3}}
[\,\pi-\epsilon(\pi\ln 3 -4{\bf C})+{\rm O}(\epsilon^2)\,],
\label{a18}\\
\partial{\cal T}_3 &=& -\frac{\Gamma(2+\epsilon)}{3}
\left[1-\frac{\pi}{3\sqrt{3}}+\epsilon
\left(1-\frac{14\pi}{3\sqrt{3}}-\frac{\pi\ln 3}{3\sqrt{3}}+
\frac{4{\bf C}}{3\sqrt{3}}\right)+{\rm O}(\epsilon^2)\right].
\label{a19}
\end{eqnarray}
{\em Sunset:}
\begin{eqnarray}
{\cal S}_0 &=& \partial{\cal S}_0=0,
\label{a20}\\
{\cal S}_1 &=&
2i\frac{\Gamma(2\epsilon)}{\epsilon}B(1-\epsilon,1-\epsilon)
B(1+\epsilon,2-4\epsilon),
\label{a21}\\
\partial{\cal S}_1 &=& -\frac{\epsilon}{2}
{\cal S}_1,
\label{a22}\\
{\cal S}_2 &=& i\frac{[\,\Gamma(\epsilon)\,]^2}%
{(1-2\epsilon)(1-\epsilon)},
\label{a23}\\
\partial{\cal S}_2 &=& -i
\frac{\Gamma(2\epsilon)B(1+\epsilon,1+\epsilon)}%
{(2-\epsilon)(1-\epsilon)},
\label{a24}\\
{\cal S}_3  &=& i\frac{[\,\Gamma(\epsilon )\,] ^2}
{1-\epsilon }\left(
\frac{1}{1-2\epsilon }+1-\frac{1}{(1+2\epsilon )(2-\epsilon )}
\right)+\frac{3}{4}i+{\rm O}(\epsilon ),\\
\partial {\cal S}_3 &=& -i\frac{1}{4\epsilon }\left[ \,\Gamma
(1+\epsilon )\, \right]^2+i\frac{3}{8}+{\rm O}(\epsilon ),
\end{eqnarray}
{\em Acorns:}
\begin{eqnarray}
{\cal A}_{10} &=& -\Gamma(2\epsilon)B(\epsilon,1-2\epsilon)
\frac{B(1-\epsilon,1-\epsilon)}{1-\epsilon},
\label{a27}\\
\partial{\cal A}_{10} &=& \frac{\epsilon}{2-\epsilon}
{\cal A}_{10},
\label{a28}\\
{\cal A}_{1m} &=& \frac{\Gamma (\epsilon) \Gamma (2\epsilon)
\Gamma (1 - \epsilon)}{1 - \epsilon}
- \frac{ \left[\,
\Gamma(\epsilon) \Gamma (1 - \epsilon)\,\right]^2}%
{(1 - \epsilon) \Gamma (2 - 2\epsilon)}
e^{i\pi\epsilon} + 2 \zeta (2) - 3
+{\rm O}(\epsilon),\\
\partial {\cal A}_{1m} &=& \frac{\Gamma (\epsilon) \Gamma (1 +
\epsilon)}{1 - \epsilon} B (1 - \epsilon, 1 - \epsilon) + 2 -
\zeta (2)+{\rm O}(\epsilon),  \\
{\cal A}_{20} &=& -\frac{\Gamma(2\epsilon)}{\epsilon}
B(1-\epsilon,1-\epsilon)B(1+\epsilon,1-2\epsilon)+2\zeta(2)
+{\rm O}(\epsilon),
\label{a29}\\
\partial{\cal A}_{20} &=&
-\frac{\Gamma(1+2\epsilon)B(1-\epsilon,1-\epsilon)}{(2+\epsilon)
\epsilon}-2+\zeta(2)+{\rm O}(\epsilon),
\label{30}\\
{\cal A}_{2m} &=& {\cal A}_4 - 2 \zeta (2)+{\rm O}(\epsilon),\\
\partial {\cal A}_{2m} &=& \partial {\cal A}_4
+ \frac{1}{3} \zeta (2) - \frac{2{\bf C}}{\sqrt{3}}
+{\rm O}(\epsilon),\\
{\cal A}_3 &=&
-\frac{\Gamma(2\epsilon)B(1-\epsilon,1-\epsilon)}%
{(1-\epsilon)\epsilon}+2\sqrt{3}{\bf C}
+{\rm O}(\epsilon),\label{a31}\\
\partial{\cal A}_3 &=& -\frac{\Gamma(1+2\epsilon)B(1-\epsilon,
1-\epsilon)}{(2-\epsilon)(1-\epsilon)\epsilon}+1+{\rm O}(\epsilon),
\label{a32}\\
{\cal A}_4 &=& -\Gamma(2\epsilon)\left[%
\frac{B(1-\epsilon,1-\epsilon)}{(1+\epsilon)\epsilon}
+2B(1-\epsilon,1-\epsilon)\left(2-\frac{\pi}{\sqrt{3}}\right)\right]
\nonumber\\
&&-\left(4-\frac{2\pi}{\sqrt{3}}+\frac{\pi\ln3}{\sqrt{3}}
-\frac{1}{2}\zeta(2)-\frac{7{\bf C}}{\sqrt{3}}\right)
+{\rm O}(\epsilon),
\label{a33}\\
\partial{\cal A}_4 &=&
\frac{\Gamma(1+2\epsilon)B(1-\epsilon,1-\epsilon)}{\epsilon}
\left(1-\frac{2\pi}{3\sqrt{3}}\right)\nonumber\\
&& + \left(3-\frac{\pi}{\sqrt{3}}+\frac{2\pi\ln3}{3\sqrt{3}}
-\frac{1}{3}\zeta(2)-\frac{8{\bf C}}{3\sqrt{3}}\right)
+{\rm O}(\epsilon),
\label{a34}\\
{\cal A}_{1Q} &=& \frac{\Gamma(2\epsilon)}{\epsilon(1-\epsilon)}
\Gamma (1 + \epsilon) \Gamma (1- \epsilon),\label{a35}\\
{\cal A}_{2Q} &=&
-\frac{\Gamma(2\epsilon)B(1-\epsilon,1-\epsilon)}%
{\epsilon(1-\epsilon)}+1-\zeta(2)
+{\rm O}(\epsilon),\label{a36}\\
{\cal A}_{4Q} &=&
-\frac{\Gamma(2\epsilon)B(1-\epsilon,1-\epsilon)}{\epsilon(1+
\epsilon)}+1+\frac{2{\bf C}}{\sqrt{3}}+{\rm O}(\epsilon),\\
\partial {\cal A}_{iQ} &=& 0,\qquad i = 1,2,4.
\label{a37}
\end{eqnarray}
{\em Auxiliary Acorns:}
\begin{eqnarray}
{\cal A}_{1m} (\hat s) &=& \Gamma (\epsilon) \left[\,
\Gamma (1 - \epsilon)\,\right]^2
\left( \frac{\Gamma(2\epsilon)}{\Gamma(2 - \epsilon)}
- \frac{\Gamma (\epsilon)(-\hat s)^{-\epsilon}}%
{(1 - \epsilon) \Gamma (2 - 2\epsilon)} \right)\nonumber\\
&& + \left( 1 + \frac{1}{\hat s} \right) {\rm Li}_2 (\hat s) +
2\left( 1 - \frac{1}{\hat s} \right) \ln (1 - \hat s)
- 3 + {\rm O}(\epsilon),\\
\frac{\partial{\cal A}_{1m}}{\partial\hat a^2} (\hat s) &=&
-\ln (-\hat s) + \left(1 - \frac{1}{\hat s} \right)
\ln (1 - \hat s) + \frac{1}{\hat s}{\rm Li}_2(\hat s)
+ {\rm O}(\epsilon),\\
\frac{\partial{\cal A}_{1m}}{\partial\hat b^2} (\hat s) &=&
-\frac{\left[\,
\Gamma (\epsilon) \Gamma (1 - \epsilon)\,\right]^2}%
{(1 - \epsilon) \Gamma (2 - 2\epsilon)} (-\hat s)^{-\epsilon}
\nonumber\\
&& +\Gamma (-\epsilon) \Gamma (1 - \epsilon)
\sum_{n=0}^\infty \frac{(n+1+\epsilon) \Gamma (n+\epsilon)
\Gamma (n + 1 + 2\epsilon)}%
{\Gamma (n+2) \Gamma(n+2-\epsilon)} \hat s^n,\\
\frac{\partial{\cal A}_{1m}}{\partial\hat d^2} (\hat s) &=&
-\frac{\left[\,
\Gamma(\epsilon) \Gamma (1 - \epsilon)\,\right]^2}%
{(1 - \epsilon) \Gamma (1 - 2\epsilon)} (-\hat s)^{-1-\epsilon}
\nonumber\\
&& -\Gamma (-\epsilon)\Gamma (1-\epsilon) \sum_{n=0}^\infty
\frac{\Gamma(n+1+\epsilon)\Gamma(n+1+2\epsilon)}%
{(n+2)\Gamma(n+2-\epsilon)\Gamma(n+1)} \hat s^n \nonumber\\
&=& -\frac{[\,\Gamma(\epsilon)\Gamma(1-\epsilon)\,]^2}%
{(1-\epsilon)\Gamma(1-2\epsilon)} (-\hat s)^{-1-\epsilon}
\nonumber\\
&& - \Gamma(-\epsilon) \Gamma (1-\epsilon)
\left[ \frac{1}{\hat s} - \frac{1}{\hat s}\left(1 -
\frac{1}{\hat s} \right) \ln (1 - \hat s) \right]
+ \frac{1}{\hat s} {\rm Li}_2(\hat s) \nonumber\\
&& + \frac{1}{\hat s} \left[ 1 + \left(1 - \frac{1}{\hat s} \right)
\ln (1 - \hat s) - 2 \left( 1 - \frac{1}{\hat s} \right)
\ln^2 (1 - \hat s ) \right] + {\rm O}(\epsilon),\\
\frac{\partial{\cal A}_{2m}}{\partial \hat a^2} (\hat s)
&=& \Gamma (1 + 2\epsilon) B (-\epsilon, 1-\epsilon)
\left[ - \left( 1 - \frac{1}{\hat s} \right) \ln ( 1 - \hat s)
\right. \nonumber\\
&& \left.+ \frac{2(4 - \hat s)}{\sqrt{4\hat s - \hat s^2}}
\arctan \frac{\hat s }{\sqrt{4\hat s - \hat s^2}} \right]
\nonumber\\
&&-\int_0^1 dy\, \left( \ln^2 (A-y) - \ln^2 A + \ln y\,
[\,
\ln A - \ln (A - y)\,] - {\rm Li}_2 \left(\frac{y}{A}\right)\right)
\nonumber\\
&& + {\rm O}(\epsilon),\qquad \mbox{where } A = 1 - \hat sy
(1 - y),\ \hat s \leq 4, \\
\frac{\partial{\cal A}_{2m}}{\partial \hat d^2} (\hat s)
&=& \frac{1}{\epsilon}\Gamma (1 + 2\epsilon) B(1 - \epsilon,
1 - \epsilon) \frac{2}{\sqrt{4\hat s - \hat s^2}}
\arctan \frac{\hat s }{\sqrt{4\hat s - \hat s^2}} \nonumber\\
&& + \int_0^1 dy\, \frac{1 - y}{A} \left[\, \ln y - \ln A
- \ln (A - y)\,\right] + {\rm O}(\epsilon)\\
\frac{\partial{\cal A}_4}{\partial\hat d^2}
&=& \Gamma(2\epsilon)B(1-\epsilon,1-\epsilon)
\frac{2\pi}{3\sqrt{3}}-\frac{\pi\ln3}{\sqrt{3}}-\zeta(2)
+\frac{7{\bf C}}{3\sqrt{3}}+{\rm O}(\epsilon).
\label{a44}
\end{eqnarray}
{\em Eye-in-the-Sky:}
\begin{eqnarray}
{\cal E}_1(\hat s) &=& -i\frac{[\,\Gamma(\epsilon)\Gamma
(1-\epsilon)\,]^2}{(1-\epsilon)\Gamma(1-2\epsilon)}
(-\hat s)^{-1-\epsilon} + i\left(\frac{1}{2}\ln(-\hat s)\right.
\nonumber\\
&&\left.-\frac{1}{2}\left(1-\frac{1}{\hat s^2}\right)\ln(1-\hat s)
+\frac{1}{2\hat s}-\frac{1}{\hat s}{\rm Li}_2(\hat s)\right) +
{\rm O}(\epsilon),\\
{\cal E}_1 &=& i\frac{[\,\Gamma(\epsilon)\Gamma(1-\epsilon)\,]^2}%
{(1-\epsilon)\Gamma(1-2\epsilon)}e^{i\pi\epsilon}
+i\left(\frac{1}{2}-\zeta(2)-i\frac{\pi}{2}\right)
+{\rm O}(\epsilon),\\
\partial{\cal E}_1 &=&
-i(1+\epsilon) \frac{[\,\Gamma(\epsilon)\Gamma(1-\epsilon)\,]^2}%
{(1-\epsilon)\Gamma(1-2\epsilon)}e^{i\pi\epsilon}
+i[\,\zeta(2)-1\,]+{\rm O}(\epsilon),\\
{\cal E}_2 &=&
i({\cal A}_{20}-{\cal A}_{1Q}), \label{a48}\\
\partial {\cal E}_2 &=& i\left( \frac{\Gamma (2+\epsilon )\Gamma
(\epsilon )}{(1-\epsilon )}B(3,-\epsilon )+2\zeta (2)-\frac{7}{2}
\right)+{\rm O}(\epsilon ),\label{a49}\\
{\cal E}_2^* &=& i\Gamma(1+2\epsilon)\frac{B(1-\epsilon,
1-\epsilon)}{1-\epsilon}B(\epsilon,1-2\epsilon),
\label{a50}\\
\partial{\cal E}_2^* &=& \frac{1+2\epsilon}{2(2-\epsilon)}{\cal
E}_2^* ,
\label{a51}\\
{\cal E}_3 &=& {\cal E}_5 + i \frac{5}{6}\zeta(2)
+{\rm O}(\epsilon),\label{a52}\\
\partial{\cal E}_3 &=&
i\Gamma(2+2\epsilon)B(1-\epsilon,1-\epsilon)
\frac{1}{\epsilon}\left(\frac{1}{3}-\frac{\pi}{9\sqrt{3}}\right)
\nonumber\\
&&-i\left(\frac{2}{3}+\frac{\pi^2}{12}-\frac{\pi\ln 3}%
{9\sqrt{3}}+\frac{7{\bf C}}{9\sqrt{3}}-
\frac{\pi}{\sqrt{3}}\right)+{\rm O}(\epsilon),
\label{a53}\\
{\cal E}_4 &=& i\frac{\Gamma(1+2\epsilon)B(1-\epsilon,1-\epsilon)}%
{(1-\epsilon)\epsilon}
+i\left(4
-\frac{4{\bf C}}{\sqrt{3}}\right)
+{\rm O}(\epsilon),\label{a54}\\
\partial{\cal E}_4 &=& i\frac{\Gamma(2+2\epsilon)B(1-\epsilon,
1-\epsilon)}{(2-\epsilon)(1-\epsilon)\epsilon}-i\frac{3}{2}
+{\rm O}(\epsilon),
\label{a55}\\
{\cal E}_5 &=& i\Gamma(2\epsilon)B(1-\epsilon,1-\epsilon)
\frac{2\pi}{3\sqrt{3}}\nonumber\\
&& +i\left(-\frac{\pi\ln3}{3\sqrt{3}}+\frac{7{\bf C}}%
{3\sqrt{3}}-\frac{\pi}{\sqrt{3}}-\frac{1}{6}\zeta(2)\right)
+{\rm O}(\epsilon),
\label{a56}\\
\partial{\cal E}_5 &=&
i\Gamma(2+2\epsilon)B(1-\epsilon,1-\epsilon)\frac{1}{\epsilon}
\left(\frac{1}{3}-\frac{\pi}{9\sqrt{3}}\right)\nonumber\\
&& -i\left(\frac{2}{3}-\frac{\pi^2}{36}-\frac{\pi\ln 3}%
{9\sqrt{3}}+\frac{31{\bf C}}{9\sqrt{3}}-\frac{\pi}{\sqrt{3}}
\right)+{\rm O}(\epsilon).
\label{a57}
\end{eqnarray}
{\em Lemons:}
\begin{eqnarray}
{\cal L}_1(\hat s) &=& \frac{1}{\hat s}
\biggl[2{\rm Li}_3(-\hat s)-2{\rm Li}_3(\hat s)
-4{\rm Li}_3(1-\hat s)-4{\rm Li}_3(1+\hat s)\nonumber\\
&&-2\ln \hat s\,{\rm Li}_2(-\hat s)
-\ln^2 \hat s\ln(1+\hat s)-\pi^2\ln(1+\hat s)
+\zeta(3)+2i\pi\ln^2(1+\hat s)\nonumber\\
&& +4{\rm Li}_3\left(\frac{1+\hat s}{1-\hat s}\right)
-4{\rm Li}_3
\left(-\frac{1+\hat s}{1-\hat s}
\right)
\nonumber\\
&& \left.+4\ln(1-\hat s)\left({\rm
Li}_2(1+\hat s) +\frac{3}{2}\zeta(2)\right)\right]+{\rm
O}(\epsilon), \label{a58}\\
{\cal L}_1 &=& -i6\zeta(3)+{\rm O}(\epsilon),
\label{a59}\\
\partial{\cal L}_1 &=& i[\,6\zeta(3)-2\zeta(2)\,]+{\rm O}
(\epsilon),
\label{a60}\\
{\cal L}_{20} &=& i[\,\zeta(2)+4\ln 2\,]+{\rm O}(\epsilon),
\label{a61}\\
\partial{\cal L}_{20} &=& i\frac{\Gamma(2+2\epsilon)B(1+\epsilon,
1-\epsilon)B(3,-2\epsilon)}{(2-\epsilon)(1-\epsilon)}
\nonumber\\
&&+i\left(-\frac{17}{54}+\frac{8}{9}\ln2+\frac{2}{3}\zeta(2)\right)
+{\rm O}(\epsilon),
\label{a62}\\
{\cal L}_{2m} &=&
i\left(4\ln2-\frac{3}{2}\zeta(3)-\frac{1}{2}\zeta(2)+
\frac{9}{4}\zeta(2)\ln2+K_2+2\pi iK'_2\right)+{\rm O}(\epsilon)
\label{a63}\\
\partial{\cal L}_{2m} &=& -{\cal L}_{2m}+i
\left(\zeta(2) - \frac{2{\bf C}}{\sqrt{3}} + i\frac{2\pi^2}%
{3\sqrt{3}}
\right)+{\rm O}(\epsilon),
\label{a64}\\
{\cal L}_3 &=& i\left(-4\ln2-\zeta(2)+6\sqrt{5}\ln
\left(\frac{\sqrt{5}+1}{2}\right)+2\ln^2\left(%
\frac{\sqrt{5}+1}{2}\right)\right)+{\rm O}(\epsilon),
\label{a65}\\
\partial{\cal L}_3 &=& i\biggl[4-4\ln2-\frac{1}{2}\zeta(2)
+8\ln^2\left(\frac{\sqrt{5}+1}{2}
\right)
\nonumber\\
&& -\frac{7\sqrt{5}}{2}
\ln\left(\frac{\sqrt{5}+1}{2}\right)
+K_3+{\rm O}(\epsilon)
\biggr],\label{a66}\\
{\cal L}_5 &=& iK_5,\label{a67}\\
\partial{\cal L}_5 &=& -i\left(K_5+\frac{1}{6}\zeta(2)
+\frac{{\bf C}}{\sqrt{3}}\right)
+{\rm O}(\epsilon).\label{a68}
\end{eqnarray}
{\em Auxiliary Lemon integrals:}
\begin{eqnarray}
K_2 &=& -\int_0^1 ds\,
\frac{4\ln s}{\sqrt{4s-s^2}}\arctan\frac{s}{\sqrt{4s-s^2}}
\nonumber\\
&&+\int_0^1 ds \int_0^1 dy \frac{s}{2-s}
\Biggl[ \frac{1-2y}{a}\ln y - \frac{1-y}{a}\ln(1-sy)
\nonumber\\
&&-\frac{1}{s}\left(\ln^2 (a-y) - \case{1}/{2}\ln^2 a
+\ln y\,\ln\frac{a}{a-y} - {\rm Li}_2
\left(\frac{a}{y}\right)\right)\Biggr],\nonumber\\
&& a=1-sy(1-y),
\end{eqnarray}
\begin{eqnarray}
K_2' &=& \int_0^1 ds\,
\frac{2}{\sqrt{4s-s^2}}\arctan\frac{s}{\sqrt{4s-s^2}},
\\
K_3 &=& \int_0^1 dx \int_0^1 dy\,
\frac{x(1-y)}{1-xy}
\Biggl[ \frac{1}{1-xy}
\left( 4x(1-x)+x^2y+\frac{x^2(1-x)^2(1-y)}{1-xy}\right)
\nonumber\\
&& \times\ln\left(1+\frac{1-xy}{x(1-x)}\right)
-3 - \frac{x(1-x)(1-y)}{1-xy}\Biggr],\\
K_5 &=& \int_0^1 dx\int_0^1 dy\,
\frac{1}{xy}\ln\left(1+\frac{b_xb_y}{2}
(1+a_x+a_y-\sqrt{(1+a_x+a_y)^2-4a_xa_y})\right),\nonumber\\
&& a_z=\frac{1-z}{z},\quad b_z=\case{1}/{2}
\left(1-\sqrt{1+\displaystyle\frac{4z}{1-z}}\right).
\end{eqnarray}

\end{flushleft}

\section{USE OF A BARNES'-TYPE REPRESENTATION TO EVALUATE\protect\\
INTEGRALS OVER FEYNMAN PARAMETERS}\label{appb}

We have found it very useful in evaluating a number of the
integrals over Feynman parameters to use a Barnes' type
representation for the binomial expansion \cite{whittaker}
\begin{equation}
\Gamma(a)(1-z)^{-a}=
\frac{1}{2\pi i}\int_{-i\infty}^{i\infty} dt\, \Gamma(t+a)\,
\Gamma(-t)\,(-z)^t,\quad |{\rm arg}(-z)|<\pi,
\end{equation}
where the contour of integration runs to the right of the poles of
$\Gamma(t+a)$, i.e., $t=-a,-a-1,\ldots$, and to the left of the poles
of $\Gamma(-t)$, i.e., $t=0,1,\ldots$. The binomial expansion of
$(1-z)^{-a}$ for $|z|<1$ (or $|z|>1$) is recovered when the
contour of integration is closed at infinity in the right half
(left half) of the complex $t$ plane. The advantage of the
integral representation is that it holds for any magnitude of
$|z|$ and any $a$.

We will illustrate the use of the Barnes' representation by
example. We first consider ${\cal A}_{1m}(\hat s)$ which is given
up to a factor $\xi^{2\epsilon}/(4\pi)^4$ by the expression in
Eq.\ (34) with $\hat a=1,\ \hat b=\hat c=\hat d=0$:
\begin{eqnarray}
{\cal A}_{1m}(\hat s) &=& -\Gamma(2\epsilon)
\int_0^1 dx\, x^{-\epsilon}(1-x)^{-\epsilon}\,
\int_0^1 dy\, y^\epsilon\, \int_0^1 dz\, z^{\epsilon-1}
\left[ -\hat s y(1-y) + \frac{yz}{x}\right]^{-2\epsilon}
\nonumber\\
&=& -\Gamma(2\epsilon) \int_0^1 dx\, x^\epsilon (1-x)^{-\epsilon}
\int_0^1 dy\, y^{-\epsilon} \int_0^1 dz\, z^{-\epsilon-1}
\left(1 - \frac{\hat s x(1-y)}{z}\right)^{-2\epsilon}\!\!.
\end{eqnarray}
We can use the Barnes' representation for the last factor to
re\,express ${\cal A}_{1m}(\hat s)$ as
\begin{eqnarray}
{\cal A}_{1m}(\hat s) &=& -\frac{1}{2\pi i}
\int_{-i\infty}^{i\infty} dt\, \Gamma(t+2\epsilon)\Gamma
(-t)(-\hat s)^t\nonumber\\
&& \times \int_0^1 dx\, x^{t+\epsilon}(1-x)^{-\epsilon}
\int_0^1 dy\, y^{-\epsilon} (1-y)^t \int_0^1 dz\,
z^{-1-\epsilon-t}.
\end{eqnarray}
The integrals on $x$ and $y$ converge and can be expressed in
terms of $\Gamma$ functions for ${\rm Re}\, t>-1,\
\epsilon\rightarrow 0+$. The final integral converges for
${\rm Re}\,t<-\epsilon$. We find that
\begin{equation}
{\cal A}_{1m}(\hat s) = \frac{[\,\Gamma(1-\epsilon)\,]^2}{2\pi i}
\int_{-i\infty}^{i\infty} dt\, \frac{\Gamma(t+2\epsilon)
\Gamma(t+\epsilon)\Gamma(-t)}%
{(t+1)\Gamma(t+2-\epsilon)}(-\hat s)^t,
\end{equation}
where the contour of integration must cross the real axis between
$-1$ and $-\epsilon$. If we take $|\hat s|<1$, we can close the
contour of integration to the right and find that
\begin{eqnarray}
{\cal A}_{1m} &=& \Gamma(\epsilon)[\,\Gamma(1-\epsilon)\,]^2
\left( \frac{\Gamma(2\epsilon)}{\Gamma(2-\epsilon)}
-\frac{\Gamma(\epsilon)(-\hat s)^{-\epsilon}}%
{(1-\epsilon)\Gamma(2-2\epsilon)}\right)\nonumber\\
&& + \sum_{n=1}^\infty \frac{\Gamma(n+2\epsilon)%
\Gamma(n+\epsilon)}%
{\Gamma(n+2-\epsilon)\Gamma(n+2)}\hat s^n.
\end{eqnarray}
The series is nonsingular for $\epsilon\rightarrow0$ and can be
summed in that limit to obtain the expression in Eq.\ (A52), or
treated more generally in terms of hypergeometric functions for
$\epsilon\not=0$.

This calculation illustrates a fairly typical result, that we can
use the Barnes' representation to at least partially factor a
multidimensional integral over Feynman parameters, and reduce the
calculation to one of contour integration.  It was essential that
the representation held in this case for arbitrary values of $\hat
sx(1-y)/z$ subject only to the constraint $|{\rm arg}(-\hat
s)|<\pi$.  This requires that $-\hat s=e^{-i\pi}\hat s$ for $\hat
s$ identified as usual with $\hat s+i\epsilon$.

In some cases, partial factorization of the integrals is
sufficient. For example, if we write the expression for
$\partial{\cal A}_{2m}/\partial\hat d^2$ as
\begin{eqnarray}
\frac{\partial{\cal A}_{2m}}{\partial\hat d^2}(\hat s)
&=& \Gamma(1+2\epsilon) \int_0^1 dx\, x^{-\epsilon}
(1-x)^{-\epsilon} \int_0^1 dy\, y^\epsilon (1-y)
\int_0^1 dz\, z^{-1+\epsilon}\nonumber\\
&&\times [1-\hat sy(1-y)]^{-1-2\epsilon}
\left(1 - \frac{yz}{1-\hat sy(1-y)}\right)^{-1-2\epsilon}
\end{eqnarray}
and apply the Barnes' representation to the last factor and
integrate over $z$, we obtain a simple series in the variable
$y/\left[ 1-\hat sy(1-y) \right]$. The singular term in $\epsilon $
can be extracted and the remainder summed in the limit
$\epsilon\rightarrow0$. The result is the expression in Eq.\ (A57).

The method can also be used to advantage to make multiple
expansions. For example, if the propagators in the expression for
${\cal E}_1(\hat s)$ are combined in an order different from that which
leads to the expressions in Eq.\ (34), one obtains the awkward
expression
\begin{eqnarray}
{\cal E}_1(\hat s) &=& i\Gamma(1+2\epsilon)
\int_0^1 dx\, x^{1+\epsilon}(1-x)^{-\epsilon}
\int_0^1 dy\,(1-y) \int_0^1 dz\, z^2(1-z)^{-2-\epsilon}
\nonumber\\
&& \times \left(1-\frac{\hat s
xyz(1-yz)}{1-z}\right)^{-1-2\epsilon}.
\end{eqnarray}
After applying the Barnes' representation to the last factor, one
obtains an expression which involves $(1-yz)^t$. This may be
factored by a second application of the Barnes' representation.
After the $x,y,$ and $z$ integrations are performed, the contour
integrals are easily evaluated, and lead to the result in Eq.\
(A59). A simpler calculation based on Eq.\ (40) gives the same
result.

Finally, it is not always necessary or even desirable to evaluate
all the integrals at an early stage even if that is possible. An
example is provided by $\partial {\cal E}_2$ which is given by the
integral
%
% equation
%
\begin{eqnarray}
\label{appb8}
	\partial {\cal E}_2 &=& i\Gamma (2+2\epsilon
	)\int_{0}^{1}dx\,x^{-\epsilon }(1-x)^{-\epsilon
	}\int_{0}^{1}dy\,y^{2+\epsilon }(1-y)^{-1-2\epsilon
	}\int_{0}^{1}dz\,z^{-1+\epsilon }(1-z)\nonumber \\
	&&\times \left( 1+\frac{yz}{x(1-y)} \right)^{-2-2\epsilon }.
\end{eqnarray}
Use of the Barnes' representation on the last term gives a
complete factorization of the integrals over the Feynman
parameters. However, the singularities at $y=1$ in the resulting
expression cause difficulties if the $y$ integration is performed
at this stage
since convergence of the $y$ and $z$ integrations require
different signs for ${\rm Re}\,t$ or $\epsilon $. However, it is
clear from the original expression that the apparent singularities
in $y$ are absent. In this case, we therefore integrate only over
$x$ and $z$.  After closing the contour of the $t$ integration in
the {\em left\/} half plane, we obtain a series in {\em
positive\/} powers of $(1-y)$. The singular term in $\epsilon $
can be extracted, the remainder of the series summed to simple
functions for $\epsilon =0$, and the final integration on $y$
performed to obtain the result in Eq.\ (A63).

\section{AN APPLICATION OF KOTIKOV'S METHOD\protect\\
 TO THE {\em LEMON\/}
GRAPHS}\label{appc}

In this Appendix we describe the application of Kotikov's
differential equation method \cite{kotikov} to the {\em Lemon}
graphs which appear in $\Pi_H^{(2)}$. The method is based on the
identity
\begin{equation}
0=i^5\int{\cal D}k{\cal D}\ell\frac{\partial}{\partial k_\mu}
(p_\mu\Delta_a\Delta_b\Delta_c\Delta_d\Delta_e),
\end{equation}
where the propagators in the {\em Lemon} graph will be denoted by
$i\Delta_a, i\Delta_b,\ldots$ in the conventions of Fig.\
\ref{fig7}. The notation $\widehat{\cal L}[\ \cdot\ ]$ will be used to
indicate the action of the {\em Lemon} operator in Eq.\ (46) on the
quantity in the bracket, i.e., multiplication by the {\em Lemon}
integrand and integration over the loop momenta $k$ and $\ell$. For
example $\widehat{\cal L}_i[1]= {\cal L}_i$ while $\widehat{\cal
L}_1[1/\Delta_a]=i{\cal A}_{1m}$.

If we choose $p_\mu=k_\mu$ in Eq.\ (C1) and perform the derivatives
we obtain the identity
\begin{equation}
0=\widehat{\cal L}
\,[2\omega-2k^2\Delta_a-2k\cdot(k+q)\Delta_b - 2k\cdot(k-\ell)
\Delta_c ],
\end{equation}
where $2\omega =4-2\epsilon $.
The scalar products can be expressed in terms of inverse propagators and
masses, e.g.,
\begin{equation}
2k\cdot q = \frac{1}{\Delta_b}-\frac{1}{\Delta_a}-q^2 - a^2 + b^2.
\end{equation}
The identity can then be rewritten as
\begin{equation}
(4-2\omega){\cal L} =
\widehat{\cal L}
\left[
\begin{array}{l}
\Delta_c\left(-\frac{1}{\Delta_a}+\frac{1}{\Delta_d}-a^2-c^2+d^2\right)\\
+\Delta_b\left(-\frac{1}{\Delta_a}+q^2-a^2-b^2\right)-2a^2\Delta_a
\end{array}
\right].
\end{equation}
If we choose $p_\mu=\ell_\mu$ in Eq.\ (C1), we obtain instead
\begin{equation}
0=\widehat{\cal L}
\left[
\begin{array}{l}
\Delta_c\left(\frac{1}{\Delta_d}-\frac{1}{\Delta_a}+c^2+d^2-a^2\right)\\
+\Delta_a\left(\frac{1}{\Delta_c}-\frac{1}{\Delta_d}+c^2-d^2-a^2
\right)\\
+\Delta_b\left(\frac{1}{\Delta_c}-\frac{1}{\Delta_a}-\frac{1}{\Delta_e}
+q^2+c^2-e^2-a^2\right)
\end{array}
\right].
\end{equation}
Further identities can be derived by replacing $\partial/\partial k_\mu$ by
$\partial/\partial\ell_\mu$ in Eq.\ (C1). These amount to interchanges of
labels
in the relations above, with $a^2\leftrightarrow d^2$ and $b^2\leftrightarrow
e^2$. Several of the {\em Lemon} graphs are symmetric under this
interchange, or under $a^2\leftrightarrow b^2$ corresponding to the interchange
of momenta $k\rightarrow -k-q,\ell\rightarrow -\ell-q$
which leads to further identities.

The result of differentiating a {\em Lemon} with respect to the square of the
incoming momentum is the identity
\begin{equation}
\frac{\partial{\cal L}}{\partial q^2}
=\frac{q_\mu}{2q^2}\frac{\partial{\cal L}}{\partial q_\mu}
=\frac{1}{2q^2}\widehat{\cal L}\,
[-2q\cdot(k+q)\Delta_b-2q\cdot(\ell+q)\Delta_e].
\end{equation}
This can be rewritten as the differential equation
\begin{equation}
\frac{\partial{\cal L}}{\partial q^2}+\frac{{\cal L}}{q^2}
=\frac{1}{2q^2}\widehat{\cal L}
\left[
\frac{\Delta_a}{\Delta_b}+(b^2-a^2-q^2)\Delta_a+
\frac{\Delta_d}{\Delta_e}+(e^2-d^2-q^2)\Delta_d\right].
\end{equation}

The key observation at this point is that the terms in the expressions above
which involve ratios of $\Delta$'s are related to derivatives of {\em Acorns}
or {\em Double bubbles}. For example, the ratio $\Delta_a/\Delta_b$ on the
right hand side of Eq.\ (C7) removes the $b$ line in a {\em Lemon}, thus
contracting two vertices, and introduces an extra $a$ propagator. The result is
$\widehat{\cal L}(\Delta_a/\Delta_b)=i\partial{\cal A}(a,c,d,e)/\partial a^2$,
i.e., $a\rightarrow a,b\rightarrow c,c\rightarrow d$, and $d\rightarrow e$ in
the {\em Acorn} graph in Fig.\ \ref{fig7}.  Terms of the form
(mass)$^2\times\Delta$ are derivatives of {\em Lemons}. The problem is to use
the identities above and the symmetries to eliminate the unwanted {\em Lemon}
terms in Eq.\ (C7) to obtain a differential equation for ${\cal L}$ with the
inhomogeneous term expressed in terms of known graphs. The results for the
cases of interest are:
\begin{eqnarray}
\frac{\partial{\cal L}_5}{\partial s}+\frac{{\cal L}_5}{s}
&=&
\frac{1}{s}\widehat{{\cal L}_5}
\left[
\frac{\Delta_a}{\Delta_c}-\frac{\Delta_a}{\Delta_d}\right]
=\frac{1}{s}\left({\cal T}_3{\cal B}_2-i\frac{\partial{\cal A}_4}%
{\partial d^2}\right),\\
\frac{\partial{\cal L}_{2m}}{\partial s}
+\frac{{\cal L}_{2m}}{s} &=&
\frac{1}{2-\hat s}\widehat{{\cal L}_{2m}}
\left[\frac{1}{s}
\left(\frac{\Delta_a}{\Delta_b}+\frac{\Delta_d}{\Delta_e}\right)+
\frac{\Delta_a}{\Delta_d}+\frac{\Delta_d}{\Delta_a}+
\frac{\Delta_a}{\Delta_c}-\frac{\Delta_d}{\Delta_c}\right]
\nonumber\\
&=&
\frac{1}{2-\hat s}\left( \frac{1}{s}
\left[ \frac{\partial{\cal A}_{1m}}{\partial a^2}+
\frac{\partial{\cal A}_{2m}}{\partial a^2}\right]+
\frac{\partial{\cal A}_{2m}}{\partial d^2} + \frac{\partial{\cal A}_{1m}}%
{\partial d^2}\right) \nonumber\\
&& - \frac{1}{m^2(2-\hat s)}({\cal T}_3{\cal B}_0+{\cal B}_2
{\cal T}_0),
\end{eqnarray}
and
\begin{eqnarray}
\frac{\partial{\cal L}_1}{\partial s}+\frac{{\cal L}_1}{s}
&=&
\frac{1}{1+\hat s}\widehat{{\cal L}_1}
\left[\frac{\Delta_a}{s\Delta_b}-
\frac{\Delta_a}{\Delta_d}+\frac{\Delta_a}{\Delta_c}\right]\nonumber\\
&=& \frac{i}{1+\hat s}
\left(\frac{1}{s}\frac{\partial{\cal A}_{1m}}{\partial b^2}-
\frac{\partial{\cal A}_{1m}}{\partial d^2}\right)+
\frac{1}{1+\hat s}{\cal T}_0{\cal B}_0.
\end{eqnarray}
All the terms on the left are finite for $\epsilon\rightarrow0$, but most of
those on the right have single or double poles.

The differential equations have general solutions of the form
\begin{equation}
{\cal L}(\hat s)=\frac{1}{\hat s}\left(t{\cal L}(t)+\int_t^{\hat s}
d\hat s\,\hat s\,f(\hat s)\right),
\end{equation}
where $f(\hat s)$ is the inhomogeneous term in the differential equation. A
careful study of the limiting behavior of the integrands for $\hat
s\rightarrow0$ with fixed $\epsilon>0$ shows that $\hat s{\cal L}(\hat
s)\rightarrow0$ for $\hat s\rightarrow0$ for the graphs which appear in
$\Pi_H^{(2)}$. (We did not use this method for the graphs which appear in
$\Pi_w^{(2)}$ which must be evaluated for $\hat s=0$.) We can therefore take
the lower limit of integration in Eq.\ (C11) as zero, and write the ${\cal
L}$'s as
\begin{equation}
{\cal L}(\hat s) = \lim_{\epsilon\rightarrow0}
\frac{1}{\hat s} \int_0^{\hat s} d\hat s\,\hat s\, f(\hat s),
\end{equation}
where an overall factor of $1/M_H^2$ has been suppressed following the
convention of Appendix A.

The homogeneous terms need to be calculated for arbitrary $\hat s$. This is
trivial for ${\cal B}_0(\hat s)$ and ${\cal T}_0(\hat s)$. ${\cal B}_2
(\hat s),{\cal T}_3(\hat s)$, and the derivatives of the {\em Acorn} graphs
with respect to internal masses were all evaluated using the method based on
Barnes' representation for the binomial expansion discussed in Appendix B and
illustrated there for ${\cal A}_{1m}(\hat s)$, and also by direct expansions in
powers of $\epsilon$ using the expression in Eqs.\ (19), (24), and (36). The
singularities in the individual terms in the $f$'s cancel explicitly, and the
problem was reduced to one of integration.

Some integrals in ${\cal L}_{2m}$ and $\partial{\cal L}_{2m}$ were evaluated
analytically. The rest, while perhaps not insuperable, were evaluated
numerically. The quantities actually needed are ${\cal L}_{2m}=
{\cal L}_{2m}(1)$ and $\partial{\cal L}_{2m}=(\partial{\cal L}_{2m}/\partial
\hat s)(1)$. The second can be evaluated immediately using Eq.\ (C12),
\begin{equation}
\partial{\cal L}\equiv \frac{\partial{\cal L}}{\partial\hat s}(1)
=-{\cal L}(1)+f(1).
\end{equation}
The results are given in Eqs.\ (A77) and (A78) \cite{kotikov2}.

${\cal L}_1$ was considerably easier to calculate, and was evaluated
analytically. Curiously,
\begin{equation}
{\cal L}_1 = {\cal L}_0 = -i6\zeta(3)
\end{equation}
where ${\cal L}_0$ is the result for all internal masses zero and $\hat s=1$
\cite{pascual}, but
\begin{equation}
\partial{\cal L}_1=\partial{\cal L}_0-i2\zeta(2).
\end{equation}
Finally, $\partial{\cal L}_5$ was evaluated using Eqs.\ (C8) and (C13) and the
result for ${\cal L}_5$ obtained using the hyperspherical method,
Eq.\ (52).

\section{A COLLECTION OF USEFUL INTEGRALS}\label{appd}

DeVoto and Duke \cite{devoto} give a useful compilation of integrals involving
logarithms and polylogarithms which occur in integrals over Feynman parameters.
Lewin \cite{dilog} is the standard reference on polylogarithms. We give here a
further collection of integrals which occured repeatedly in the calculations of
the finite terms in the self-energy functions. The first few are elementary;
the remainder are not.

{\em Integrals involving\/} $X=x(1-x)$:
\begin{eqnarray}
&& \int_0^1dx\,\frac{1}{1-X}=\frac{2\pi}{3\sqrt{3}},\\
&& \int_0^1dx\,\frac{1}{(1-X)^2}=\frac{2}{3}+\frac{4\pi}{9\sqrt{3}},\\
&& \int_0^1dx\,\ln(1-X)=-2+\frac{\pi}{\sqrt{3}},\\
&& \int_0^1dx\,\ln^2(1-X)=8-\frac{4\pi}{\sqrt{3}}+
\frac{2\pi\ln3}{\sqrt{3}}-\frac{8{\bf C}}{\sqrt{3}},\\
&& \int_0^1dx\,x\,\ln(1-X)=\frac{\pi}{2\sqrt{3}}-1,\\
&& \int_0^1 dx\,x^2\ln(1-X)=-\frac{1}{18},\\
&& \int_0^1dx\,\ln x\,\ln(1-X)=4-\frac{\pi}{\sqrt{3}}-\sqrt{3}{\bf C}-
 \frac{\pi^2}{36},\\
&& \int_0^1dx\,\frac{\ln(1-X)}{1-X}=\frac{2\pi\ln3}{\sqrt{3}}
-\frac{8{\bf C}}{3\sqrt{3}},\\
&& \int_0^1 dx\,\frac{\ln(1-X)}{x}=\frac{-\pi^2}{18},\\
&& \int_0^1dx\,\left(\frac{\ln(1-X)}{x^2}+\frac{1}{x}\right)=
-1+\frac{\pi}{\sqrt{3}},\\
&& \int_0^1dx\,\frac{\ln x}{1-X}=\frac{-2{\bf C}}{\sqrt{3}},\\
&& \int_0^1 dx\,\frac{x\ln x}{1-X}=\frac{-{\bf C}}{\sqrt{3}}
+\frac{\pi^2}{36},\\
&& \int_0^1dx\,\frac{\ln x}{(1-X)^2}=\frac{-\pi}{3\sqrt{3}}-
\frac{4{\bf C}}{3\sqrt{3}},\\
&& \int_0^1dx\, \frac{x\ln x}{(1-X)^2}=\frac{-2{\bf C}}{3\sqrt{3}},\\
&& \int_0^1dx\,\frac{\ln(1+X)}{x}=2\ln^2\left(\frac{\sqrt{5}-1}{2}\right).
\end{eqnarray}
Note that many further useful identities can be obtained by substituting
$x\rightarrow (1-x)$, and also by using the identity
\begin{equation}
\int_0^1dx\, x\,f(X)=\frac{1}{2}\int_0^1dx\,f(X).
\end{equation}

{\em Integrals involving the dilogarithm\/} \cite{dilog} ${\rm Li}_2(z)$:
\begin{eqnarray}
\int_0^1 dx\,{\rm Li}_2(x) &=& \zeta(2)-1,\\
\int_0^1dx\,\ln x\,{\rm Li}_2(x) &=& 3-2\zeta(2),\\
\int_0^1dx\, \ln(1-x){\rm Li}_2(x) &=& 3-2\zeta(3)-\zeta(2),\\
\int_0^1dx\,{\rm Li}_2\left(\frac{-1}{x}\right) &=&
-\frac{1}{2}\zeta(2)-2\ln2,\\
\int_0^1 dx{\rm Li}_2(X) &=& \frac{\pi^2}{18}+
\frac{2\pi}{\sqrt{3}}-4,\\
\int_0^1dx\,{\rm Li}_2\left(\frac{1-X}{-X}\right) &=&
-2\sqrt{3}{\bf C},\\
\int_0^1dx\,\left(\frac{{\rm Li}_2(X)}{x^2}-\frac{1}{x}\right)
&=& -\frac{\pi}{\sqrt{3}}-\frac{\pi^2}{9}.
\end{eqnarray}

\begin{figure}
	\caption{The one-loop self-energy graphs needed in the calculation
of $\Pi_H^{(1)}$ and $\Pi_w^{(1)}$. The heavy lines represent Higgs
bosons, and the thin lines, Goldstone bosons. Tadpole graphs cancel
exactly, and are omitted.
\label{fig1}}
\end{figure}
\begin{figure}
	\caption{The two-loop self-energy graphs needed in the calculation of
$\Pi_w^{(2)}$. The weights with which individual graphs
appear in $\Pi _w^0$ can be read off from Eq.\ (11).
\label{fig2}}
\end{figure}
\begin{figure}
	\caption{The two-loop self-energy graphs needed in the calculation of
$\Pi_H^{(2)}$. The weights with which individual graphs appear in
$\Pi^0 _H$ can be read off from Eq.\ (12).
\label{fig3}}
\end{figure}
\begin{figure}
	\caption{The scattering graphs through two loops which involve only the
dimension-four quartic couplings and give finite or logarithmically
divergent contributions to the renormalized scattering amplitude for
$s,-t,-u\rightarrow\infty$. Particle masses are irrelevant in this
limit so all internal lines may be taken as massless.
\label{fig4}}
\end{figure}
\begin{figure}
	\caption{Examples of scattering graphs which contain dimension-three cubic
couplings and give vanishing contributions to the renormalized
scattering amplitude for $s,-t,-u\rightarrow\infty$. For example,
the scattering {\em Eye} graph ${\cal E}_1(\hat s)$ with the
topology of (d) given in Eq.\ (A59) vanishes as $s^{-1}\ln^2 s$ for
$s\rightarrow\infty$ after renormalization.
\label{fig5}}
\end{figure}
\begin{figure}
	\caption{The conventions used for the particle masses and momenta in the
one-loop
self-energy graphs.
\label{fig6}}
\end{figure}
\begin{figure}
	\caption{The conventions used for the particle masses and momenta in the
two-loop
self-energy graphs.
\label{fig7}}
\end{figure}


\begin{references}
\bibitem{fixing}
Most of the derivations of the equivalence theorem [1,3,17--19]
are based on variations of the Feynman--'t\,Hooft gauge-fixing
condition
that
$\langle ik_\mu W_a^\mu+M\phi_a\rangle=0$ for any
physical matrix element, where $W_a^\mu,\phi_a$ are the vector
boson and scalar fields in the theory. With the observation that
$k^\mu/M=\xi_L^\mu+{\rm O}(M/k^0)$, this becomes a relation
between amplitudes which involve longitudinally-polarized vector
bosons, and amplitudes which involve the corresponding Goldstone
scalars. The constant of proportionality depends on the
renormalization conditions for the fields $W_a^\mu$ and $\phi_a$
\cite{bs,he}.
H. Veltman \cite{hvelt} shows by detailed analysis that the
replacement of $k^\mu/M$ by
the longitudinal polarization vector $\xi_L^\mu$,
which appears as a weak point of earlier derivations of
the equivalence theorem, is legitimate and causes no problem
despite extensive gauge cancellations. (See, however, the
comments in \cite{he} on the renormalization scheme used in
\cite{hvelt} in which $C\not=1$ in higher order.)
\bibitem{old32}
More precisely,
$2\rightarrow2$
diagrams which involve dimension-three
interactions in scattering processes are suppressed at a fixed
scattering angle by powers of $M_H^2/s$. The partial-wave
scattering amplitudes with which we will be concerned will
similarly be suppressed by powers of $M_H^2/s$; the extra factors
of $\ln(s/M_H^2)$ which appear do not cause problems.
\end{references}
\end{document}